\documentclass[aps,pra,amsmath,amssymb]{revtex4-1}
\usepackage{hyperref}
\usepackage{graphicx}
\usepackage[normalem]{ulem}
\usepackage{bbm}
\usepackage{bm}
\usepackage[caption=false]{subfig}
\usepackage{color}
\usepackage{dsfont}

\DeclareMathOperator{\tr}{tr}

\newcommand{\phid}{\phi^\dag}
\newcommand{\cphi}{\varphi}

\newcommand{\ctheta}{\vartheta}
\newcommand{\PHI}{\mathbf{\Phi}}
\newcommand{\MR}{\mathbf{R}}
\newcommand{\MG}{\mathbf{G}}
\newcommand{\MGamma}{\bm{\Gamma}}
\newcommand{\Q}{\mathbf{q}}
\newcommand{\X}{\mathbf{x}}
\newcommand{\vP}{\mathbf{p}}
\newcommand{\vGamma}{\dot{\Gamma}}
\newcommand{\vPHI}{\dot{\PHI}}
\newcommand{\vphi}{\dot{\phi}}

\newcommand{\vsigma}{\dot{\sigma}}
\newcommand{\vctheta}{\dot{\ctheta}}
\newcommand{\vrho}{\dot{\rho}}
\newcommand{\vu}[1]{\dot{u}_{#1}}
\newcommand{\vn}[1]{\dot{n}_{#1}}

\newcommand{\vZctheta}{\dot{Z}_\ctheta}
\newcommand{\vYm}{\dot{Y}_m}

\newcommand{\vZm}{\dot{Z}_m}

\newcommand{\rhok}[1]{\rho_{#1,k}}
\newcommand{\uk}[1]{u_{#1,k}}

\newcommand{\bk}{b_k}
\newcommand{\vbk}{\dot{b}_k}

\newcommand{\ddX}{\mathrm{d}^d\X}

\newcommand{\ddQ}{\mathrm{d}^d\Q}

\begin{document}

\title{Application  of the functional renormalization group to Bose gases: from linear to hydrodynamic fluctuations}

\author{Felipe Isaule}
\author{Michael C. Birse}
\author{Niels R. Walet}
\affiliation{Theoretical Physics Division, School of Physics and Astronomy, The University of Manchester, Manchester, M13 9PL, United Kingdom}

\date{\today}

\begin{abstract}

We study weakly interacting Bose gases using the functional renormalization group with a hydrodynamic effective action. 
We use a scale-dependent parametrization of the boson fields that interpolates between a Cartesian representation at high momenta and an amplitude-phase one for low momenta. We apply this to Bose gases in two and three dimensions near the superfluid phase transition where they can be described by statistical O(2) models. 
We are able to give consistent physical descriptions of the infrared regime in both two and three dimensions.
In particular, and in contrast to previous studies using the functional renormalization group, 
we find a stable superfluid phase at finite temperatures in two dimensions. We compare our results
for the superfluid and boson densities with Monte-Carlo simulations, and we find they are in reasonable agreement.
\end{abstract}


\maketitle

\section{Introduction}
\label{sec:Intro}

Weakly interacting Bose gases have long been of theoretical interest for studying the multitude of phenomena related to Bose-Einstein condensation and superfluidity \cite{Bogoliubov1947,Anderson1995,Davis1995} (see Ref.~\cite{Pitaevskii2016} for a review). The phase transition between the normal and the superfluid  state is particularly interesting, not least because this can be studied, both theoretically and experimentally, in  three, two and even one dimension. The character of these transitions depends sensitively on the dimensionality. 

Especially with the modern development of techniques for cooling gases of alkali atoms, these phenomena can be explored in great detail experimentally, see Refs.~\cite{Pethick2008,Chin2010,Pitaevskii2016} for reviews.  One particularly useful aspect of the study of atomic gases is the fact that the interaction between the atoms can be tuned using Feshbach resonances, allowing the behavior of these systems to be explored for a range of interaction parameters. Of particular interest for this work are studies of low-dimensional systems \cite{Gorlitz2001,Hadzibabic2006,Clade2009}, since these show significantly different features from their more straightforward three-dimensional counterparts.

This is closely linked to the fact that a consistent theoretical description of such systems is more difficult for lower dimensions.
The mean-field approximation gives a qualitative description of superfluidity in three dimensional systems but, for quantitative results and to be able to describe the normal-to-superfluid phase transition, effects of fluctuations need to be included, see, e.g., Ref.~\cite{Andersen2004}. In particular, it is important to include the Goldstone modes of the system, which represent excitations of the phase of the condensate.

The effects of these fluctuations increase as the dimensionality decreases, and they can even suppress condensation \cite{AlKhawaja2002,AlKhawaja2002a,Posazhennikova2006}. For example, in one dimension, there is no Bose-Einstein condensate (BEC) at \emph{any} temperature. In two dimensions, condensation can occur only at zero temperature, as required by the Coleman-Mermin-Wagner theorem \cite{Mermin1966}. Nonetheless, two-dimensional systems can still display superfluid behavior, even though there is no long-range order. In those systems the phase transition occurs through the Berenzinskii-Kosterlitz-Thouless (BKT) mechanism, where vortices form bound pairs \cite{Berezinskivi1971,Kosterlitz1973}.

The traditional way to make a detailed theoretical description for such problems is through the use of well-known many-body techniques. These generally rely on a diagrammatic representation of many-body perturbation theory, where a subset of contributions is summed to infinite order. 
One difficulty faced by such approaches is that the ungapped propagators for the Goldstone modes lead to infrared (IR) divergences in many of these contributions. However there are strong cancellations between these divergences that follow from Ward identities for the spontaneously broken U(1) symmetry \cite{1975JETPL..21....1N,Dupuis2011}. These cancellations are lost if an expansion of, say, a self-energy is truncated at finite order. (See, for example, Refs.~\cite{Pistolesi2004,Stoof2014}).

This problem affects, in particular, calculations using the most straightforward representation for the fields: the linear or Cartesian representation. In the case of a Bose gas described by a complex field, this treats the real and imaginary parts of field as independent variables with an O(2) symmetry. In the broken phase, the longitudinal component of the field describes fluctuations in the magnitude of the condensate, while the transverse part describes the Goldstone modes.
The strong coupling between the longitudinal and Goldstone modes in this representation is responsible for the IR-divergent terms in quantities like the one-loop self energy of the longitudinal field.

Similar issues arise in other areas of physics, in particular, when a linear sigma model is used to describe broken chiral symmetry in hadron physics (see, for example, \cite{Gasiorowicz1969}). There they can be avoided by working with a nonlinear sigma model \cite{Weinberg1967}, as in chiral perturbation theory \cite{Weinberg1979}. In those effective theories, 
the Goldstone modes are only weakly coupled at low momenta and so large but canceling contributions do not appear.

For the nonrelativistic Bose gas we are concerned with here, the corresponding effective theory is well-known to be the hydrodynamic one introduced by Popov \cite{Popov1987}. This employs a polar or amplitude-phase (AP) representation for the fields, where the Goldstone field has only derivative interactions. As a result, the IR divergences of the Cartesian version are absent and, at least at low momenta, these 
interactions can be treated in perturbation theory. In this representation, the Ward identities can be satisfied without delicate cancellations. On the other hand, loop integrals over the Goldstone modes are divergent in the ultraviolet (UV), which led Popov to impose a rather arbitrary cutoff. However, these divergences can be renormalized by adding higher-derivative interactions to the theory. In that case, Popov's Lagrangian should be regarded as just the lowest-order piece in the expansion of an effective field theory \cite{Liu1998}. The couplings in such a theory do not have simple connections to the interactions between the particles. For example in the context of hadron physics, their values are fixed using either phenomenological input or full simulations of quantum chromodynamics.

A powerful alternative approach is the functional renormalization group (FRG) \cite{Wetterich1993,Berges2002,Kopietz2010}. This is a nonperturbative technique based on the evaluation of the full effective action. A scale-dependent regulator is introduced, and one follows the  flow of the couplings in this action  with the regulator scale. The flow starts from a bare classical action where all fluctuations, whether quantum or thermal, are suppressed by the regulator. In the limit where the regulator scale 
$k$ is taken to zero, the full effective action (the generator of one-particle irreducible Green's functions) is recovered. In practical applications, the expansion of the action is truncated to a finite number of terms. As a result the action in the physical limit will depend on the choices made in setting up the flow equations. These include the form of the regulator, and ways of optimizing this choice have been developed \cite{Litim2001, Pawlowski2007}.

There are many versions of the FRG; in this work we use the one introduced by Wetterich \cite{Wetterich1993,Berges2002} for the average effective action, which is based on the Legendre transform of the logarithm of the partition function. The exact flow equation for this effective action has a one-loop structure. However, in practice, a truncation scheme needs to be used in order to solve for the RG flow. In this work we employ a derivative expansion; other choices can be found in the literature \cite{Blaizot2005,Blaizot2006,Machado2010}.  Truncations based on derivative expansions have been widely used in applications of the FRG and, despite the approximations introduced, they have been successful in describing phase transitions and critical exponents in a variety of systems \cite{Berges2002,Kopietz2010}.

In particular, with fields in the Cartesian representation, the FRG has had some success in describing the weakly-interacting Bose gas and related O($N$) models in three dimensions. In the critical regime, it yields  results that are in good agreement with Monte-Carlo simulations. Away from that region, it has also been used to investigate bulk thermodynamic properties \cite{Berges2002,Canet2003,Blaizot2005,Floerchinger2008,
Floerchinger2009b,Eichler2009,Dupuis2011,Rancon2012a}.
As shown by Dupuis \cite{Dupuis2009,Dupuis2011}, one appealing aspect of this FRG approach is that, despite the potential IR divergences of loop diagrams as the physical limit is approached, it does respect the relevant Ward identity \cite{1975JETPL..21....1N}, giving a vanishing anomalous self-energy.

This self-energy, and the resulting IR-divergent propagator for the longitudinal mode, arise from a momentum-independent interaction that vanishes in the physical limit. This treatment thus fails to capture the interaction between the amplitude fluctuations of the condensate that is present in Popov's effective theory. It also omits the leading interaction between the Goldstone modes, which is of second order in momenta. These limitations may be remedied by the inclusion of higher-order terms in the derivative expansion of the action. However, in current calculations applying this scheme to three-dimensional Bose gases, the flow needs to be stopped at a finite scale $k$, before reaching the physical effective action, since the IR divergence occurs at a very small but finite $k$ \cite{Wetterich2008,Floerchinger2008,Dupuis2011}.
Fortunately, various bulk quantities such as the condensate density are insensitive to this scale, and converge
to their physical values before it is reached \cite{Dupuis2011}.

Such IR problems are more acute in two dimensions, where the FRG has not been as successful. As in three dimensions, straightforward applications of the FRG in the Cartesian representation to the O(2) model \cite{Grater1995,Gersdorff2001,Dupuis2011} and the Bose gas \cite{Floerchinger2009c,Dupuis2011,Rancon2012} show a similar breakdown of the flow at small scales. Nevertheless, these authors suggest that the approach may describe universality in the behavior near the breakdown scale, reproducing, for example, the critical anomalous dimension. However, this is based on a line of pseudo-fixed points at finite running scales. For a given initial condition, the flow reaches a reasonable value for the anomalous dimension only for a single value of $k$, before becoming unstable. 
Below that scale, the flow is always driven to the normal (non-superfluid) phase at any finite temperature, implying a finite correlation length, in contradiction to the BKT physics. A stable superfluid phase has been found in this framework, but only at a cost: either fine-tuning of the regulator \cite{Jakubczyk2014} or unphysical neglect of the longitudinal mode \cite{Jakubczyk2017}. Again, it has been argued that the instability is an artifact of the truncation and that adding higher-order terms will lead to the correct flow \cite{Berges2002,Jakubczyk2017}, but this has not yet been demonstrated in detail.

An intriguing feature of these results from the Cartesian representation is that, even though a critical point cannot be located unambiguously, the FRG seems able to recover aspects of vortex physics without explicit reference to vortices. This deserves further exploration, for instance using different field representations. 

These issues with the Cartesian version of the FRG for  two-dimensional systems have motivated recent papers implementing versions of the FRG in AP representations, to a lattice system in Ref.~\cite{Krieg2017} and to continuum boson fields in Ref.~\cite{Defenu2017}. The work of Defenu \textit{et al.} \cite{Defenu2017} shows that, with fields in the AP representation,  the FRG is able to reproduce a physical superfluid phase at lowest order of the derivative expansion, without having to stop the evolution at a finite scale. However, using the AP representation for all momentum scales runs into problems in the UV. In contrast to the momentum-independent bare interactions between the particles, the derivative couplings of the Goldstone modes grow with momentum. As mentioned above, this makes it nontrivial to match a theory in the AP form onto the underlying one. Instead, 
the authors of Ref.~\cite{Defenu2017} simply subtract the contributions of free fields, for which the path integral is Gaussian. This corresponds to an uncontrolled UV renormalization. 
Apart from the lack of consistency in the treatment of fluctuations, this approach is specifically designed for the region close the phase transition only.

In the present work, we develop a hybrid version of the FRG, using a basis for the boson fields that smoothly switches between the Cartesian and AP representations, which we refer to as an ``interpolating representation". This is based on the ideas of Pawlowski as presented by Lamprecht in Ref.~\cite{Lamprecht2007} 
who proposed a convenient form for this interpolation. It allows us to start 
in the Cartesian representation for high regulator scales and then at lower scales, where phase fluctuations become increasingly dominant, to smoothly  switch to the AP form. 

In this first application of the method, we study an $O(2)$ model in both two and three dimensions. This correspond to a classical statistical approach (the classical field approximation)  for weakly-interacting 
Bose gases close to their phase transitions \cite{Baym2001}. 
We examine the effect on the RG flow of switching representations and check that we recover the correct IR behavior in the physical limit. By using the same approach in both two and three dimensions, we can study the dependence on the dimensionality of the system. In particular, we focus on obtaining a superfluid phase in two dimensions within a simple truncation in order to demonstrate that the interpolating representation simplifies the description of this state.

We use the FRG to compute the superfluid and boson densities, comparing our results with Monte-Carlo simulations of Refs.~\cite{Prokofev2002,Prokofev2004}. The interpolator proposed by Lamprecht and Pawlowski \cite{Lamprecht2007} contains two parameters one of which controls the scale at which the switch between representations is made and the other determines how readily the switch is made. We examine the dependence on both of these. 
We find  stable results if the switch is made at a momentum scale where the longitudinal mode starts behaving very differently from the phase (Goldstone) modes. This corresponds to the hydrodynamic or ``healing" scale \cite{Pitaevskii2016}.

Our article is organized as follows. In section \ref{sec:Action} we define our microscopic model, the FRG formalism and the truncation scheme used. Then in section \ref{sec:BrokenPhase}, we give details of the Cartesian and AP representations we use, and outline the distinction between the condensate and quasi-condensate densities in the superfluid phase. We then give details of the interpolator we use to switch between these representations, and present the resulting flow equations. Finally, in section \ref{sec:Results} we present our results for two and three dimensions, comparing them with 
the results from Monte-Carlo simulations.

\section{Effective action}
\label{sec:Action}

We consider a system of bosons at finite temperature weakly interacting through a short-range repulsive potential. This  has a phase transition to a superfluid phase. For temperatures close to the critical one and considering only low-momentum modes, we can work in the ``classical field approximation"
\cite{Baym2001}, where the time dependence of the fields is neglected. 
This is valid for momenta $|\Q|\ll\lambda_{DB}^{-1}$, where 
$\lambda_{DB}=\sqrt{2\pi/mT}$ is the thermal wavelength. 
This description is equivalent to an $O(2)$ model.  This is a very useful starting point to test our approach. 
Expressed in terms of the complex boson field, $\phi$, the bare action takes the form
\begin{equation}
\mathcal{S}[\PHI]=-\,\frac{1}{ T} \int\ddX \left[\frac{1}{2m}\nabla\phid\nabla\phi -\mu_0\phid\phi+\frac{g}{2}(\phid\phi)^2\right].
\label{eq:S}
\end{equation}
Here $T$ is the temperature, $\mu_0$ the chemical potential and $g$ the strength of 
a repulsive contact interaction.
Here and in the following we express all mentioned quantities in units where $\hbar=k_B=1$. We have 
also introduced $\PHI=(\phi,\phid)$ to denote a vector containing the field and its
conjugate.

Starting from the action (\ref{eq:S}), we use the FRG to obtain a renormalization group equation for the Legendre-transformed effective action which depends on classical fields $\PHI^{cl}$
\cite{Wetterich1993,Berges2002}. In this approach, the flow is driven by a regulator $\MR(\Q,k)$, which is added
 to the theory to suppress  fluctuations for momenta $|\Q|<k$. 
The resulting effective action $\Gamma_k[\PHI^{cl}]$ runs with the scale $k$ of the regulator. For a large enough UV scale, $k=\Lambda$, fluctuations are  completely suppressed and 
$\Gamma_\Lambda[\PHI^{\text{cl}}]$ can be taken to be the bare action, 
\begin{equation}
\Gamma_\Lambda[\PHI^{\text{cl}}]=\mathcal{S}[\PHI^{\text{cl}}].
\end{equation} 
In contrast, for $k\rightarrow 0$ all fluctuations are taken into account and 
$\Gamma_0[\PHI^{\text{cl}}]$ corresponds to the effective action which describes the physics of the interacting system. This functional 
 is the generator of the one-particle irreducible Green's functions for the system. The FRG constructs this action by following the flow with respect 
to $k$, starting from the bare action at the UV scale $k=\Lambda$. Note that, 
from now on, we drop the superscript ``$\text{cl}$" and will take $\PHI$ to refer to 
the classical fields.

\begin{figure}
    \subfloat[]{\includegraphics[scale=1.8]{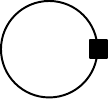}}
    \hspace{1cm}
	\subfloat[]{\includegraphics[scale=1.8]{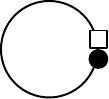}}
    \caption{\label{fig:FlowEq} Diagrammatic representations of the terms in 
the flow equations (\ref{eq:WetterichEq}) and (\ref{eq:PawlowskEq}). 
The filled square denotes $\partial_k \MR$, the empty square $\MR$  and the filled circle  $\vPHI^{(1)}$.}    
\end{figure}

The evolution of the action $\Gamma_k[\PHI]$ as a function of $k$ is governed 
by the Wetterich equation \cite{Wetterich1993,Berges2002},
\begin{equation}
 \partial_k \Gamma=\frac{1}{2}\tr\left[\partial_k \MR \left(\MGamma^{(2)}-\MR\right)^{-1}\right],
 \label{eq:WetterichEq}
\end{equation}
where $\MGamma^{(2)}$ is the matrix of second functional derivatives with respect to the fields $\PHI$. The driving term of this equation has a one-loop structure, 
which can be represented by the diagram in Fig.\ \ref{fig:FlowEq}(a), 
if we identify $G=(\MGamma^{(2)}-\MR)^{-1}$ as the propagator. More generally, the fields can be allowed to explicitly depend on $k$. In that case, 
the flow equation is modified to \cite{Pawlowski2007},
\begin{align}
\partial_{k}\Gamma+\vPHI\cdot\frac{\delta \Gamma}{\delta \PHI}=\frac{1}{2}\tr\left[\partial_k \MR(\MGamma^{(2)}-\MR)^{-1}\right] +\tr\left[\vPHI^{(1)} \MR(\MGamma_k^{(2)}-\MR)^{-1}\right].
\label{eq:PawlowskEq}
\end{align}
Here $\partial_{k}\Gamma$ represents the $k$ derivative for constant fields,
$\vPHI=\partial_k \PHI$ is the $k$ derivative of the fields, 
and $\vPHI^{(1)}$ is the matrix of the first functional derivatives of $\vPHI$ 
with respect to the fields. The diagrammatic representation of the additional 
term is shown in Fig.\ \ref{fig:FlowEq}(b).

In general the flow equation, either Eq.~(\ref{eq:WetterichEq}) or (\ref{eq:PawlowskEq}), 
is a functional differential equation that can not be solved directly.
A simplifying approximation needs to be made, and a
 common approach is  to use an Ansatz based on a gradient expansion 
of $\Gamma$, truncated to a small number of scale-dependent terms. This reduces the functional FRG equations to a
set of coupled ordinary differential equations for the scale-dependent couplings, which can be integrated numerically using standard methods.

In this work, we use the following Ansatz for the O(2) model:
\begin{equation}
 \Gamma[\PHI]=-\frac{1}{T} \int \ddX\left[\frac{Z_m}{2m}\nabla\phid\nabla\phi+\frac{Y_m}{8m}\nabla\rho\nabla\rho+U(\rho,\mu) \right],
 \label{eq:Gammak}
\end{equation}
where $Z_m$ and $Y_m$ are $k$-dependent mass-renormalization factors. The function
 $U(\rho,\mu)$ is the effective potential expressed in terms of 
$\rho=\phid\phi$. We expand this potential to quartic order in the fields 
around its ($k$-dependent) minimum $\rho_0$, which corresponds to the vacuum expectation value of $\rho$. 

We choose to work at fixed chemical potential, taking the physical chemical 
potential $\mu_0$ to be a $k$-independent parameter. This means that the boson 
density $n_0$ depends on $k$. An alternative would be to run the evolution 
at constant density, in which case $\mu_0$ would depend on $k$ 
as in Ref.\ \cite{Birse2005}.
Following the approach of Floerchinger and Wetterich \cite{Floerchinger2008}, 
we expand $U(\rho,\mu)$ around $\mu_0$. This allows us to evaluate derivatives 
with respect to the chemical potential $\mu$, for instance to determine 
the boson density via
\begin{equation}
n_0=-\partial_\mu U|_{\mu=\mu_0}. 
\end{equation}
We thus keep terms up to first order in the shift in the chemical potential 
$\mu-\mu_0$ from its physical value, which results in the parametrization of the potential 
\begin{multline}
 U(\rho,\mu)=u_0+u_1(\rho-\rho_0)+\frac{u_2}{2}(\rho-\rho_0)^2-n_0(\mu-\mu_0)-n_1(\mu-\mu_0)(\rho-\rho_0)-\frac{n_2}{2}(\mu-\mu_0)(\rho-\rho_0)^2,
 \label{eq:Ueff}
\end{multline}
where the coefficients $u_i$ and $n_i$ all run with $k$. 

There is a redundancy in specifying both the parameters $u_1$ and $\rho_0$. We keep both so that the same form of the potential can deal with a transition between the broken and symmetric phase.
If, during the flow, the system is the symmetric phase, we must have $\rho_0=0$ and we
allow $u_1$ to run. On the other hand, in the phase where the $U(1)$ symmetry is 
broken, we follow the evolution of the position of the minimum of the potential, $\rho_0$, and we
set $u_1=0$. A non-zero value of $\rho_0$ signals the occurrence of superfluidity and the value of $Z_m \rho_0$ at $k=0$ enables us to extract the physical superfluid density. The detailed physical interpretation of $\rho_0$ will be discussed below in subsection 
\ref{sec:BrokenPhase,sub:Representations}. Since we evolve at fixed chemical potential, we determine the physical boson density at the end of the evolution
from the value of $n_0$ at $k=0$.

We start the flow at a high scale $k=\Lambda\ll \lambda_{DB}^{-1}$ (discussed in detail in subsection \ref{sec:BrokenPhase,sub:bk}), where we impose initial conditions 
that match the running action Eq.~(\ref{eq:Gammak}) to the bare action 
Eq.~(\ref{eq:S}):
\begin{align}
\rho_{0,\Lambda}=n_{0,\Lambda}=\frac{\mu_0}{g}\Theta(\mu_0),\quad u_{1,\Lambda}=-\mu_0\Theta(-\mu_0),\quad u_{2,\Lambda}=g, \nonumber \\
\quad Z_{m,\Lambda}=1, \quad Y_{m,\Lambda}=0, \quad n_{1,\Lambda}=1, \quad n_{2,\Lambda}=0,
\end{align}
where $\Theta$ denotes the unit step function.
If the evolution starts with the system in the symmetric phase, 
we find that it remains in that phase throughout the evolution, which is largely trivial.
Therefore we always start with the system in the broken phase, with $\mu_0>0$ at the scale $\Lambda$. 
In cases where the physical state of the system is in the normal phase, 
$\rho_0$ reaches zero at some finite scale $k_s$, and we then continue the flow using the symmetric parametrization of the potential.

\section{Evolution in the broken phase}
\label{sec:BrokenPhase}
We now introduce the interpolating representation used 
for the fields in the broken phase and we present in detail the resulting flow 
equations. The corresponding equations for the symmetric phase can be
obtained straightforwardly. They are given in
Appendix \ref{app:SymmetricPhase}. 

\subsection{Field representations}
\label{sec:BrokenPhase,sub:Representations}

Since the  key issue we explore in this work is the effect of the choice of field
representation on the RG evolution in a phase with a condensate, we concentrate on this part of problem. 
In this case the effective potential has a minimum at a nonzero value 
of $\rho=\phid\phi=\rho_0$. Choosing the corresponding expectation value
of the field $\phi$ to be real and positive, we can decompose the fields as 
$\phi=\sqrt{\rho_0}+\sigma+i\pi$.  Here $\sigma$ describes the fluctuations 
of the longitudinal (or ``Higgs") mode and $\pi$ the fluctuations of the 
gapless Goldstone mode. This is the Cartesian representation of the field, which has been 
widely used in FRG studies of phase transitions in $O(N)$ models and Bose 
gases \cite{Delamotte2004,Dupuis2007,Floerchinger2008,Jakubczyk2017}. 

An alternative is the AP representation, 
as in the hydrodynamic effective theory \cite{Popov1987}. This representation ensures that
low-momentum Goldstone modes interact only weakly and, as a result, symmetry 
constraints are satisfied without requiring delicate cancellations 
of large or even divergent contributions. In this work we use the representation 
\begin{equation}
\phi=(\sqrt{\rho_0}+\sigma)e^{i\ctheta/\sqrt{\rho_0}},
\end{equation}
where $\sigma$ now describes the amplitude (radial) fluctuations and $\ctheta$ 
the Goldstone (phase) fluctuations (note that in this case the variable $\ctheta$ has periodicity $2\pi\sqrt{\rho_0}$). Other versions can be found in the literature; 
for example, Defenu \textit{et al.}~\cite{Defenu2017} use 
$\phi=\sqrt{\rho_0+\sigma}e^{i\ctheta}$ in their recent application of the FRG to 
two-dimensional systems.

An important point to note is that the parameter $\rho_0$ plays a very different role in each of these representations. In the Cartesian case, 
$\sqrt{\rho_0}$ is equal to the expectation value of the longitudinal component of 
the field. If this is nonzero, the vacuum is not invariant under the symmetry transformations 
and thus $\rho_0$ is an order parameter for the system. For a Bose gas,
this  is simply the condensate density, denoted by $\rho_c$. 
On the other hand, in the AP representation, $\sqrt{\rho_0}$ is equal to 
the expectation value of the radial field, which is invariant under the symmetry. Popov originally referred this quantity $\rho_0$ as the ``bare condensate" \cite{Popov1987}, but it is now commonly referred as the ``quasi-condensate" density $\rho_q$ \cite{Kagan1987}.

The radial part of the field, defined in terms of $\rho=\phi^\dagger\phi$, is a U(1) invariant. Even if this is held fixed, its phase can still fluctuate, suppressing the long-range order. As a result, the value of the quasi-condensate is always larger than the condensate, and significantly so if there are large fluctuations in the phase of the field.
Since it is a U(1) invariant, the quasi-condensate density is not an order parameter
for a broken symmetry. Indeed it is possible for a system to have a 
finite $\rho_q$ even though $\rho_c$ is zero and the symmetry is unbroken. 
In particular, this is what gives rise to the BKT phase transition, which is driven by vortex structures in the phase field.

To illustrate these features in more detail, consider the long-distance behavior of the off-diagonal correlation function,
\begin{equation}
G_n(\X)=\langle\phid(\X)\phid(0)\rangle.
\label{eq:GnX}
\end{equation}
For a system with long-range order (LRO) this tends to the condensate density as ${|\X|\to\infty}$,
\begin{equation}G_n(\X)\underset{x\rightarrow\infty}{\sim}\rho_c.
\end{equation}
This describes the behavior in three dimensions below the critical temperature $T_c$, where the order parameter is finite. In contrast, in two dimensions $\rho_c$ can be nonzero only at $T=0$.
Above the critical temperature the symmetry is unbroken, the condensate vanishes, and thus $G_n(\X)$ decays exponentially.
For such a system described in the Cartesian representation, it is straightforward to see that 
$G_n(\X)$ approaches a finite value at long distances, and this 
leads to the relation $\rho_c=\rho_0$. Therefore, in FRG calculations that use this representation, the physical limit of $\rho_0$ as $k\rightarrow 0$ is identified with the condensate density $\rho_c$. These calculations find that, in three dimensions, $\rho_0$ saturates at a finite value for $k\to0$, as expected \cite{Blaizot2005,Floerchinger2008}.

In cases where the system is superfluid but the symmetry is unbroken, such as in the two-dimensional Bose gas at finite temperatures below the BKT transition, $G_n(\X)$ decays as a power law rather than as an exponential:
\begin{equation}
G_n(\X) \underset{x\rightarrow\infty}{\sim} \rho_q(|\X|/\xi)^{-\eta},
\end{equation}
where $\eta$ is the anomalous dimension and $\xi$ the correlation length.
Here $\rho_q$ is the quasi-condensate density just discussed, which can be interpreted as a local condensate with a fluctuating phase. (More formal definitions and modern discussions of $\rho_q$ can be found in Refs.~\cite{Prokofev2002,Prokofev2004,AlKhawaja2002,Capogrosso-Sansone2010,Stoof2014}.)
Since $G_n(\X)$ vanishes as $|\X|\to\infty$, the condensate density is zero and there is no order parameter, even though $\rho_q$ is finite. This behavior is known as ``quasi-long-range order" (QLRO).
In the FRG, a power-law behavior of $G_n(\X)$ means that $\rho_c$ is also expected to decay as a power law for small $k$, $\rho_c\sim k^{\eta}$, tending to zero as $k\rightarrow 0$. However, as mentioned in the introduction, this has not been fully achieved with the truncations used in calculations so far \cite{Berges2002,Jakubczyk2017}.

In the AP representation, the correlation function takes the form
\begin{align}
G_n(\X)=&\left\langle(\sqrt{\rho_{0}}+\sigma(\X))
(\sqrt{\rho_{0}}+\sigma(0))
e^{i(\ctheta(\X)-\ctheta(0))/\sqrt{\rho_0}}\right\rangle \nonumber\\
=&\rho_{0}\left\langle e^{i(\ctheta(\X)-\ctheta(0))/\sqrt{\rho_{0}}}\right\rangle,
\label{eq:GnX_AP}
\end{align}
where the long-distance behavior is determined by the average over phase 
fluctuations in the exponential. These fluctuations mean that the actual
condensate density is smaller than $\rho_0$, which should now be identified as the quasi-condensate density $\rho_q$.

\subsection{Interpolating representation}

The AP representation is appropriate for describing low-momentum 
Goldstone modes \cite{Prokofev2004} as it directly implements the
symmetry constraints on the interactions of those modes. 
In particular it is crucial for applications of the FRG to systems 
where large fluctuations in those modes mean there is only QLRO, such as 
the two-dimensional Bose gas. 

In contrast, at high momenta  the gap for longitudinal modes is less important,
and longitudinal and transverse fluctuations are expected to be of similar sizes.
In the high cutoff-scale regime of the FRG, the Cartesian basis is thus 
the appropriate one. Indeed it is used to define the bare action that the FRG flow starts from.

In order to apply the FRG to systems where fluctuations have a significant 
effect on the condensate or indeed destroy the LRO, we need to switch between 
the Cartesian and AP representations. A convenient tool for doing this is
the scale-dependent field representation proposed by
Lamprecht and Pawlowski \cite{Lamprecht2007}. This provides a $k$-dependent 
basis for the fields that interpolates smoothly between the two representations. In subsection \ref{sec:BrokenPhase,sub:bk} we will define in detail the regions where each representation is used.

In the interpolating representation, $k$-dependent fields $\sigma$ and $\ctheta$ 
are defined in terms of the original fields $\phi$ by \cite{Lamprecht2007}
\begin{equation}
\phi=(\sigma+\bk)e^{i\ctheta/\bk}-(\bk-\sqrt{\rho_0}).
\label{eq:InterpFields}
\end{equation}
This function $\bk$ should tend to $+\infty$ as $k\to\infty$ so that 
the fields become
\begin{equation}
\phi(\X) = (\sqrt{\rho_0}+\sigma(\X))+i\ctheta(\X),
\end{equation}
and we recover the Cartesian representation, with $\sigma$ representing the longitudinal fluctuations and $\ctheta$ the Goldstone fluctuations.
On the other hand,  $\bk$ should tend to $\sqrt{\rho_0}$ in the physical 
limit, giving the AP representation
\begin{equation}
\phi = (\sigma+\sqrt{\rho_0})e^{i \ctheta/\sqrt{\rho_0}}.
\end{equation}
In this limit, the field $\sigma$ represents the amplitude fluctuations 
and $\ctheta$ the phase fluctuations.
The specific form of the interpolating function $\bk$ that we use is 
discussed below in Sec.~\ref{sec:BrokenPhase,sub:bk}. 

In terms of the $k$-dependent fields the density $\rho=\phid \phi$ takes the form
\begin{equation}
\rho=\bk^2\left[A_k^2(\sigma)+B_k^2-2A_k(\sigma)B_k\cos(\ctheta/\bk)\right]
\label{eq:rhoInterp}
\end{equation}
where
\begin{equation}
A_k(\sigma)=\left(1+\frac{\sigma}{\bk}\right), \, B_k=\left(1-\frac{\sqrt{\rho_0}}{\bk}\right).
\label{eq:AkBk}
\end{equation}
At the minimum of the potential, $\sigma=\ctheta=0$, the density is
given by $\rho=\rho_0$ for all values of $k$.

The fields $\sigma$ and $\ctheta$ depend on the scale $k$, both explicitly through
$\bk$ and implicitly through the running parameter $\rho_0$. 
For this reason we need to employ the modified flow equation (\ref{eq:PawlowskEq}).
On the left-hand side this contains the $k$-derivatives of the fields $\vPHI=(\vsigma,\vctheta)$.
These can be found by differentiating Eq.~(\ref{eq:InterpFields}), using the
$k$-independence of the original fields ($\vphi=\vphi^\dag=0$) to obtain
\begin{align}
\vsigma=&-\vbk-\left(\frac{1}{2\sqrt{\rho_0}}\vrho_0-\vbk\right)\cos(\ctheta/\bk), \label{eq:vsigma}\\
\vctheta=&\frac{\vbk}{\bk}\ctheta+\frac{\bk}{\sigma+\bk}\left(\frac{1}{2\sqrt{\rho_0}}\vrho_0-\vbk\right)\sin(\ctheta/\bk) \label{eq:vctheta}.
\end{align}
At the minimum of the potential, $\sigma=\ctheta=0$, these become 
\begin{equation}
\vsigma\big|_{\sigma=\ctheta=0}=-\frac{1}{2\sqrt{\rho_0}}\vrho_0, \qquad \vctheta\big|_{\sigma=\ctheta=0}=0.
\end{equation}
In the final term on the right-hand side of Eq.\ (\ref{eq:PawlowskEq}), we need $\vPHI^{(1)}$ which is given by
\begin{equation}
\vPHI^{(1)}=\frac{\delta \vPHI}{\delta \PHI}=\begin{pmatrix}
 0& -\frac{\bk}{(\bk+\sigma)^2}C_k\sin(\ctheta/\bk)\\
\frac{1}{\bk}C_k\sin(\ctheta/\bk) & \frac{\vbk}{\bk}+\frac{1}{\bk+\sigma}C_k\cos(\ctheta/\bk)
 \end{pmatrix},
 \label{eq:vPHI1int}
\end{equation}
where
\begin{equation}
C_k=\left(\frac{\vrho_0}{2\sqrt{\rho_0}}-\vbk\right).
\end{equation}

\subsection{Ansatz and flow equations}
By inserting the definition (\ref{eq:InterpFields}) into  Eq.\ (\ref{eq:Gammak}) we obtain a parametrization of $\Gamma$ for the broken phase in terms of the interpolating fields. It reads
\begin{widetext}
\begin{multline}
\Gamma[\PHI]=-\frac{1}{T} \int \ddX\Bigg[\frac{Z_\ctheta}{2m}A^2_k(\sigma)(\nabla\ctheta)^2+\frac{Z_\sigma(\ctheta)}{2m}(\nabla\sigma)^2+\frac{Y_m}{2m}\bk^2\\ 
\times\Bigg(\frac{\sigma}{\bk}\left(\frac{\sigma}{\bk}+2\big(1-B_k\cos(\ctheta/\bk)\big)\right)(\nabla\sigma)^2+A^2_k(\sigma)B^2_k\sin^2(\ctheta/\bk)(\nabla\ctheta)^2\\
+2\Big(A_k(\sigma)-B_k\cos(\ctheta/\bk)\Big)A_k(\sigma)B_k\cos(\ctheta/\bk)\nabla\sigma\nabla\ctheta \Bigg)+U(\rho,\mu)\Bigg],
\label{eq:Gammabk}
\end{multline}
\end{widetext}
where $\rho$ is given by Eq.\ (\ref{eq:rhoInterp})  and 
\begin{equation}
Z_\sigma(\ctheta)=Z_\ctheta+Y_m\bk^2\big(1-B_k\cos(\ctheta/\bk)\big)^2.
\end{equation}
When evaluated at the potential minimum, this takes the form $Z_\sigma=Z_\ctheta+Y_m\rho_0$.  Depending on which limit of the interpolating basis is used, we identify $Z_\ctheta$ and $Z_\sigma$ as the mass renormalizations of the Goldstone/phase and longitudinal/amplitude modes, respectively. In  the Cartesian representation  $Z_\ctheta$ is usually denoted as $Z_\pi$. The effective potential $U$ is defined in Eq.\ (\ref{eq:Ueff}), where we take $u_1=0$ since we are working in the broken phase.

It is easy to prove that in the limit when $\bk\to\infty$  we recover the Cartesian Ansatz that can be found elsewhere \cite{Jakubczyk2017}. Similarly,  when $\bk=\sqrt{\rho_0}$ we recover the structure of the hydrodynamic actions used in other works, for example Ref.\ \cite{Popov1987}.

The propagator evaluated at $\sigma=\ctheta=0$ is given by
\begin{align}
G_{\sigma\sigma}(\Q)=&\frac{-1}{Z_\sigma\Q^2/m-2n_1\delta\mu+4(u_2-n_2\delta\mu)\rho_0+R_\sigma(\Q)}, \nonumber \\
G_{\ctheta\ctheta}(\Q)=&\frac{-1}{Z_\ctheta\Q^2/m-2n_1 B_k \delta\mu +R_\ctheta(\Q)}, \label{eq:G_b}
\end{align}
where $\delta\mu=\mu-\mu_0$ and $B_k$ is defined in Eq.\ (\ref{eq:AkBk}). The off-diagonal terms vanish, $G_{\ctheta\sigma}(\Q)=G_{\sigma\ctheta}(\Q)=0$. For calculational simplicity we use the optimized regulator \cite{Litim2001a}
\begin{equation}
R_{\cphi}(\Q)=\frac{Z_{\cphi}}{m}(k^2-\Q^2)\Theta(k^2-\Q^2),
\label{eq:Rlitim}
\end{equation}
where $\cphi=\sigma,\ctheta$, $Z_\sigma=Z_\ctheta+\rho_0Y_m$ and $\Theta(x)$ is the unit step-function. The use of different renormalization factors $Z_\cphi$ in the regulators for the longitudinal and transverse fields reflects the fact that these describe fluctuations about a vacuum with spontaneously broken symmetry. We include these factors in order to be able to solve analytically the momentum integrals and thus simplify the numerical computations. We aim to explore different choices of regulator in future work.

\begin{figure}
  \includegraphics[scale=1.5]{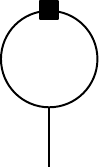}\\ 
 \includegraphics[scale=1.5]{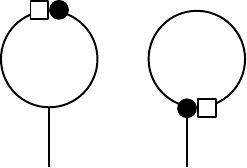}
 \caption{\label{fig:FlowGamma1_b}Diagrammatic representation of the terms contributing to $\vGamma^{(1)}$ in the broken phase. Filled squares denote $\partial_k \bm{R}$, blank squares denote $\bm{R}$ and circles denote $\vPHI^{(1)}$.}
\end{figure}

The flow equations for the $k$-dependent couplings and factors are extracted from field derivatives of Eq.\ (\ref{eq:PawlowskEq}). At the level of truncation used in this work, Eqs.\ (\ref{eq:Gammak},\ref{eq:Ueff}), they are: 
\begin{align}
2u_2\sqrt{\rho_0}\vrho_0=&\,\vGamma^{(1)}_{\sigma}\Big|_{\rho_0,\mu_0},\nonumber\\
-4\rho_0\vu2+2u_2\vrho_0=&\,\vGamma^{(2)}_{\sigma\sigma}\Big|_{\rho_0,\mu_0}, \nonumber\\
\vn0-n_1\vrho_0=&\,\partial_\mu\vGamma\Big|_{\rho_0,\mu_0}, \nonumber\\
2\sqrt{\rho_0}\vn1-2n_2\sqrt{\rho_0}\vrho_0=&\,\partial_\mu\left(\vGamma^{(1)}_{\sigma}\right)\Big|_{\rho_0,\mu_0}, \nonumber\\
4\rho_0\vn2+2\vn1-2n_2\vrho_0=&\,\partial_\mu\left(\vGamma^{(2)}_{\sigma\sigma}\right)\Big|_{\rho_0,\mu_0}, \nonumber\\
-\frac{\vZctheta}{m}=&\,\partial_{\vP^2}\left(\vGamma^{(2)}_{\ctheta\ctheta}\right)\Big|_{\rho_0,\mu_0,\vP=0}, \nonumber\\
-\frac{\rho_0\vYm}{m}-\frac{\vZctheta}{m}=&\,\partial_{\vP^2}\left(\vGamma^{(2)}_{\sigma\sigma}\right)\Big|_{\rho_0,\mu_0,\vP=0}, \label{eq:FlowEqs_b}
\end{align}
The terms on the left-hand sides come from taking derivatives for both terms on the left-hand-side of Eq.\ (\ref{eq:PawlowskEq}) and evaluating them at $\rho=\rho_0$ and $\mu=\mu_0$,  using Eqs.\ (\ref{eq:vsigma},\ref{eq:vctheta}). Similarly, the terms on the right-hand sides denote derivatives of both terms on the right-hand-side of Eq.\ (\ref{eq:PawlowskEq}). The $\sigma$ and $\ctheta$ subscripts denote  derivatives with respect to those fields. The diagrammatic representation of $\vGamma^{(1)}$ is shown in Fig.\ \ref{fig:FlowGamma1_b} and of $\vGamma^{(2)}$ is shown in Fig.\ \ref{fig:FlowGamma2_b}. The analytical expressions for all the diagrams are given in Appendix \ref{app:DiagramsBP}. The upper diagrams in both figures arises from the first term on the right-hand side of Eq.\ (\ref{eq:PawlowskEq}), while the bottom diagrams originate from the second term. The vertices between external legs and propagators correspond to the usual functional derivatives of the action in Eq.\ (\ref{eq:Gammabk}). On the other hand, vertices between external legs and $\vPHI^{(1)}$ (black circles), correspond to functional derivatives of $\vPHI^{(1)}$ in Eq.\ (\ref{eq:vPHI1int}). Because of the trigonometric functions in both the Ansatz for $\Gamma$ and in $\vPHI$,  all these diagrams contribute.  In this work, we have used \texttt{Wolfram Mathematica} \cite{Mathematica} to compute the functional derivatives, manipulate the diagrams and analytically carry out the momentum integrals  using the regulator in Eq.\ (\ref{eq:Rlitim}). 

To construct the flow equations of $Z_\ctheta$ and $Y_m$, the external legs in the corresponding diagrams must carry an external momentum $\vP$. The resulting driving terms are then expanded to the appropriate order around $\vP=0$ to determine the evolution of the mass renormalizations.  
The evolution of $u_2$ and $n_2$ is driven by $\Gamma^{(2)}_{\sigma\sigma}$, which follows from the identification of $2u_2 \rho_0$ as the mass of the longitudinal/amplitude mode. This is equivalent to differentiating  the flow equation for $U$  twice with respect to $\rho$ and then evaluating at $\rho=\rho_0$. Similarly, since $Y_m$ is tied to $Z_\sigma$, its evolution is also extracted from $\Gamma^{(2)}_{\sigma\sigma}$ .

\begin{figure}
  \includegraphics[scale=1.5]{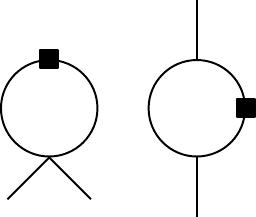}\\ 
 \includegraphics[scale=1.5]{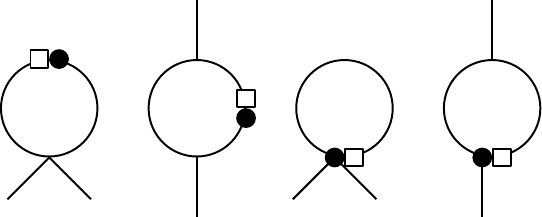}
 \caption{\label{fig:FlowGamma2_b}Diagrammatic representation of the terms contributing to $\vGamma^{(2)}$ in the broken phase. Filled squares denote $\partial_k \bm{R}$, blank squares denote $\bm{R}$ and circles denote $\vPHI^{(1)}$.}
\end{figure}

\subsection{Choice of $\bk$}
\label{sec:BrokenPhase,sub:bk}

In Popov's hydrodynamic approach \cite{Popov1987},  the high-momentum modes are described using the Cartesian representation, and the low-momentum modes in the AP representation. These two regimes are separated by a scale $k_0$ which, while not completely arbitrary, is not uniquely defined \cite{Popov1987,Dupuis2009}. To avoid a similar ambiguity in the FRG, we need to identify the scale at which $b_k$ switches the flow between the two representations. We should also check that there is only a weak dependence of observables on the precise value of this scale.
The Cartesian representation should be used in the high-momentum regime where both longitudinal and Goldstone fluctuations are important, that is, when both propagators  $G_{\sigma\sigma}$ and $G_{\ctheta\ctheta}$ in Eq.\ (\ref{eq:G_b}) are comparable. This ensures that we recover the bare action in the UV. On the other hand, the AP representation should be used in the low-momentum regime where the phase fluctuations of the Goldstone mode dominate over the amplitude fluctuations, that is when $G_{\ctheta\ctheta}\gg G_{\sigma\sigma}$. These two regimes can be distinguished in the FRG flow
by the dimensionless quantity \cite{Wetterich2008}
\begin{equation}
w_k=\frac{Z_\sigma k^2/2m}{2u_2\rho_0}.
\end{equation}
For cutoffs at high momenta, interactions between the fluctuations are suppressed and the path integral over them is approximately Gaussian. We therefore refer to this as the ``Gaussian" regime. It
corresponds to $w\gg 1$, and the low-momentum (or Goldstone) regime to $w\ll 1$. The evolution starts from UV scales $k$ deep in the Gaussian regime, and it ends in the Goldstone regime unless $\rho_0$ reaches zero during the evolution and the system goes to the symmetric phase.

By examining the behavior of the longitudinal/amplitude propagator, we can see that  $G_{\sigma\sigma}$ is dominated by the kinetic term in the Gaussian regime, whereas it is dominated by  $2u_2\rho_0$ in the Goldstone regime which suppresses the corresponding fluctuations. In contrast, the propagator $G_{\ctheta\ctheta}$ is not gapped. In a dynamical system
this would lead to a phonon-like energy spectrum  (linear in momentum), rather than particle-like (quadratic in momentum). This change in the spectrum at low momenta is closely related to the onset of superfluidity \cite{Andersen2004,Posazhennikova2006}.
As we will show later, the use of the AP representation at low momenta enables us to obtain a gapped propagator for the amplitude modes in two dimensions. Moreover, we can identify a characteristic scale $k_h$ that separates the two regimes, given by the value of $k$ where $w=1$:
\begin{equation}
k_h^2=4mu_{2}(k_h)\rho_{0}(k_h).
\label{eq:kh}
\end{equation}
We shall refer to $k_h$ as the \textit{healing scale}, in analogy to the similarly-defined physical healing length which  sets a relevant scale for superfluid phenomena \cite{Posazhennikova2006,Pitaevskii2016}. 

In order to use the AP representation in the Goldstone regime and the Cartesian representation in the Gaussian regime we use an interpolator of the form proposed by Lamprecht and Pawlowski \cite{Lamprecht2007},
\begin{equation}
\bk=\sqrt{\rho_0}\left[1+(\alpha \, w_k)^\nu\right],
\label{eq:bk}
\end{equation}
where the parameter $\alpha$ allows us to check the dependence of our results on the switching scale, and the parameter $\nu$ controls how fast the transition between the two representations is made. As just discussed, the transition should be made around the healing scale $k_h$, so $\alpha$ should be of the order of one. In the following section we show results for $\alpha=1$. We have checked that observables like the superfluid density are insensitive to variations of $\alpha$ by a factor of two (see Appendix \ref{app:alpha} for more details).  There is no similar argument to choose $\nu$. In section \ref{sec:Results} we show results for different values of $\nu$ in order to determine the best choice.

\subsection{Superfluid density}
 
From the discussion in subsection \ref{sec:BrokenPhase,sub:Representations}, it is clear that by using this interpolating representation with running fields, the parameter $\rho_0$ will change from being the scale-dependent condensate density $\rho_c$ in the Gaussian regime to the quasi-condensate density $\rho_q$ in the Goldstone regime. However, the observable quantity of interest is a third density: the superfluid density $\rho_s$. 

Here we make the obvious remark that condensation and superfluidity  are related but not identical concepts, which is most evident in systems with QLRO. Condensation  refers to the macroscopic occupation of the same quantum state and is reflected in the breaking of a global symmetry, whereas superfluidity refers to the property of particles to flow without friction.  Thus, in general, these densities are different, $\rho_s \neq \rho_c$ (for more discussion see Ref.~\cite{Yukalov2016}). Similarly, although the concept of quasi-condensate is more closely related to superfluidity, in general its density is also different.

The superfluid density can be defined from the stiffness with respect to phase changes, that is, from the coefficient of the kinetic term in the action governing the phase fluctuations at low momenta \cite{Popov1987} 
\begin{equation}
\Gamma_{\rm kin}[\cphi]=\int \ddX\, \frac{\rho_s}{2m}(\nabla \cphi)^2,
\end{equation}
where $\cphi$ is the phase field normalized to have periodicity $2\pi$.
We find that, independent of the representation used, the scale-dependent superfluid density is given by \cite{Dupuis2007}
\begin{equation}
\rho_s=Z_\ctheta \rho_0,
\label{eq:scd}
\end{equation}
This becomes the physical superfluid density in the limit $k\to 0$. In the AP representation, which corresponds to the limit $\bk=\sqrt{\rho_0}$ of our interpolator, the superfluid density is related to the quasi-condensate by $\rho_s=Z_\ctheta \rho_q$.

\section{Results}
\label{sec:Results}

In this section we present  results for the O(2) model in two and three dimensions. These are obtaining using $T=10$, $m=1$
and the initial condition $u_{2,\Lambda}=g=10^{-3}$. The value of the chemical potential is varied in order to study the behavior of the system around the phase transition.

The normal-to-superfluid phase transition point depends on the strength of the interaction.  Additionally, in the classical field approximation the UV scale $\Lambda$ is arbitrary, and different choices of $\Lambda$ move the transition point. However, the universal features of the weakly-interacting Bose gas around the phase transition can be studied by using the dimensionless functions \cite{Capogrosso-Sansone2010}
\begin{equation}
f_s=\rho_s/(m^d T^2 g^{d-2})^\frac{1}{4-d},
\end{equation}
which corresponds to the dimensionless superfluid density, and
\begin{equation}
\lambda=(n_0-n_c)/(m^dT^2g^2)^\frac{1}{4-d},
\end{equation}
which corresponds to the dimensionless density profile that indicates how the boson density deviates from its value at the phase transition. These are functions only of the dimensionless control parameter
\begin{equation}
X=\frac{\mu_0-\mu_c}{(m^dT^2g^2)^\frac{1}{4-d}},
\end{equation}
where $n_c$ and $\mu_c$ are the critical boson density and critical chemical potential, respectively. Depending on the value of $\mu_0$, the system will be in the superfluid phase ($X>0$), where the flow remains in the broken phase, or in the normal phase ($X<0$), where the flows reaches the symmetric phase at some finite scale. The value of $\mu_c$ that result in the phase-transition defines $X=0$.
 We use $\Lambda=1$ as the initial UV scale, and we choose values of $\mu_0$ which ensures that we start in the Gaussian regime. We will compare results for $f_s(X)$ and $\lambda(X)$ from our approach with those of Monte-Carlo simulations  from Refs.\ \cite{Prokofev2002,Prokofev2004}.

\subsection{Three dimensions}

\begin{figure}[t!]
\centering
\includegraphics[width=.5\textwidth]{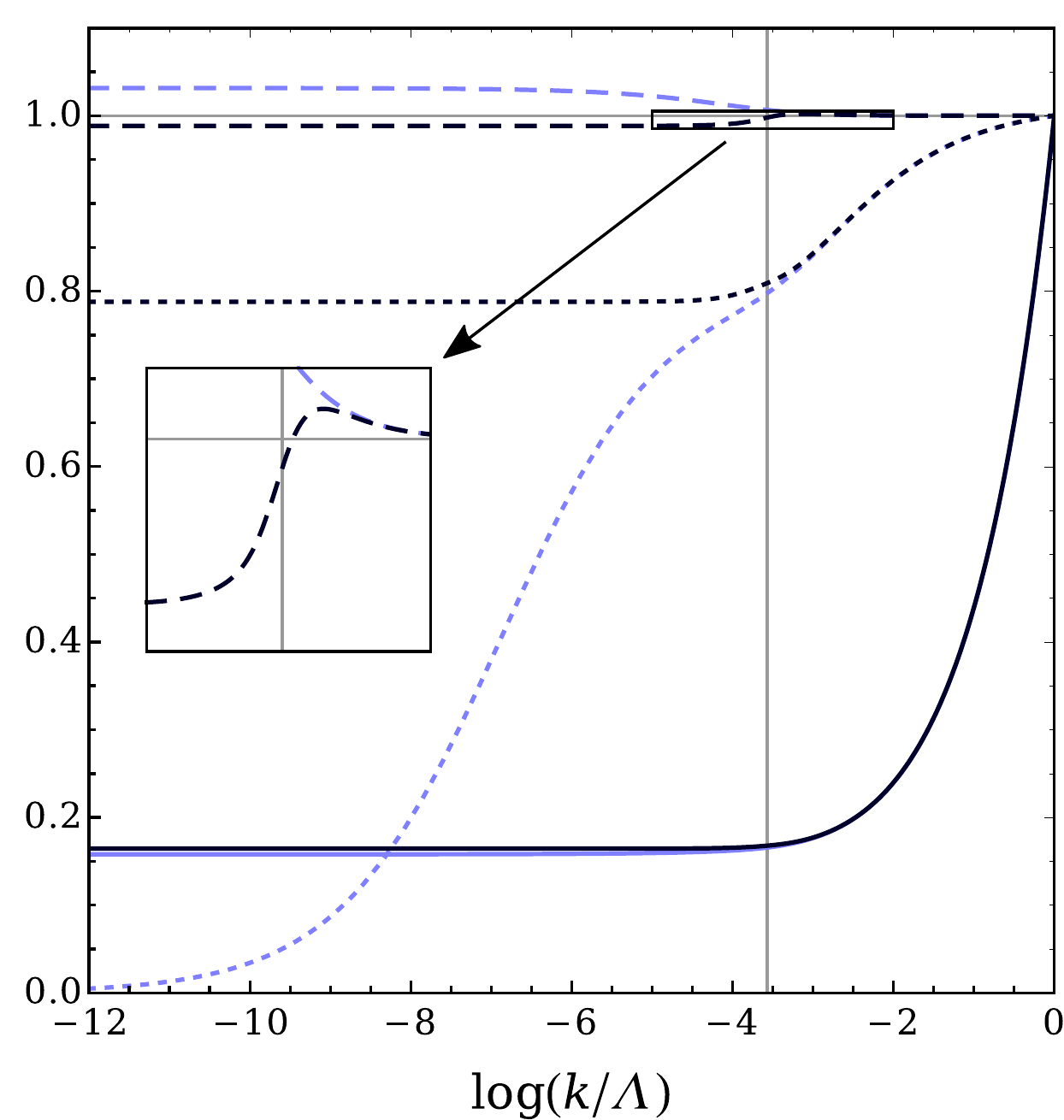}
\caption{\label{fig:flow3D} Flow of $\rho_0$ (solid lines), $u_2$ (dotted lines) and $Z_\ctheta$ (dashed lines) as a function of $\log(k/\Lambda)$ in three dimensions for $\mu_0=1.5\times10^{-3}$. The values of both $\rho_0$ and $u_2$ have been scaled by their initial values at $k=\Lambda$.  The light blue lines are the flow obtained using the Cartesian representation  and the dark blue lines are obtained using the interpolating basis with $\nu=3$. The vertical line corresponds to the scale $k_h$ in Eq.\ (\ref{eq:kh}).}
\end{figure}

Fig.\ \ref{fig:flow3D} shows an example of a typical evolution of $\rho_0$, $u_2$ and $Z_\ctheta$ in the superfluid phase as a function of $k$. We compare the evolution using the interpolating basis with the ones obtained using the Cartesian representation  as in previous FRG studies. We can see that initially, for $k\gg k_h$ where $k_h$ is the scale defined in Eq.\ (\ref{eq:kh}), both flows coincide since the interpolating basis is in its Cartesian limit. Then as $k$ approaches $k_h$ the flows start to differ. The flow of $\rho_0$ is not significantly affected, showing only a small difference in the Goldstone regime.
In contrast, the flow of $u_2$ is quite different in the two representations. In the Cartesian case it is driven to zero linearly with $k$, as has been seen in other FRG treatments \cite{Dupuis2011}, while with the interpolating basis it converges to a finite value.
$Z_\ctheta$ remains around one at the beginning of the flow (its bare value in the UV), and starts to deviate around $k_h$. With the Cartesian representation $Z_\ctheta$ increases as we lower $k$, converging to a physical value greater than one. This arises from the fact that the superfluid density $\rho_s=Z_\ctheta \rho_0$ is larger than the condensate density $\rho_c$ \cite{AlKhawaja2002,AlKhawaja2002a}, which equals  $\rho_0$ in the Cartesian representation.  On the other hand, with the interpolating basis $Z_\ctheta$ 
initially increases in the Gaussian regime, then starts to decrease around $k_h$, and finally converges to a value smaller than one.
This behavior is as expected for the interpolation since the superfluid density is smaller than the quasi-condensate density $\rho_q$, which equals $\rho_0$ in the AP limit. Thus, with the interpolation we can see that during the flow $\rho_0$ changes from being the scale-dependent condensate density $\rho_c$ to the quasi-condensate density $\rho_q$, as expected. 

\begin{figure}[t!]
\centering
 \subfloat[]{\includegraphics[width=.48\textwidth]{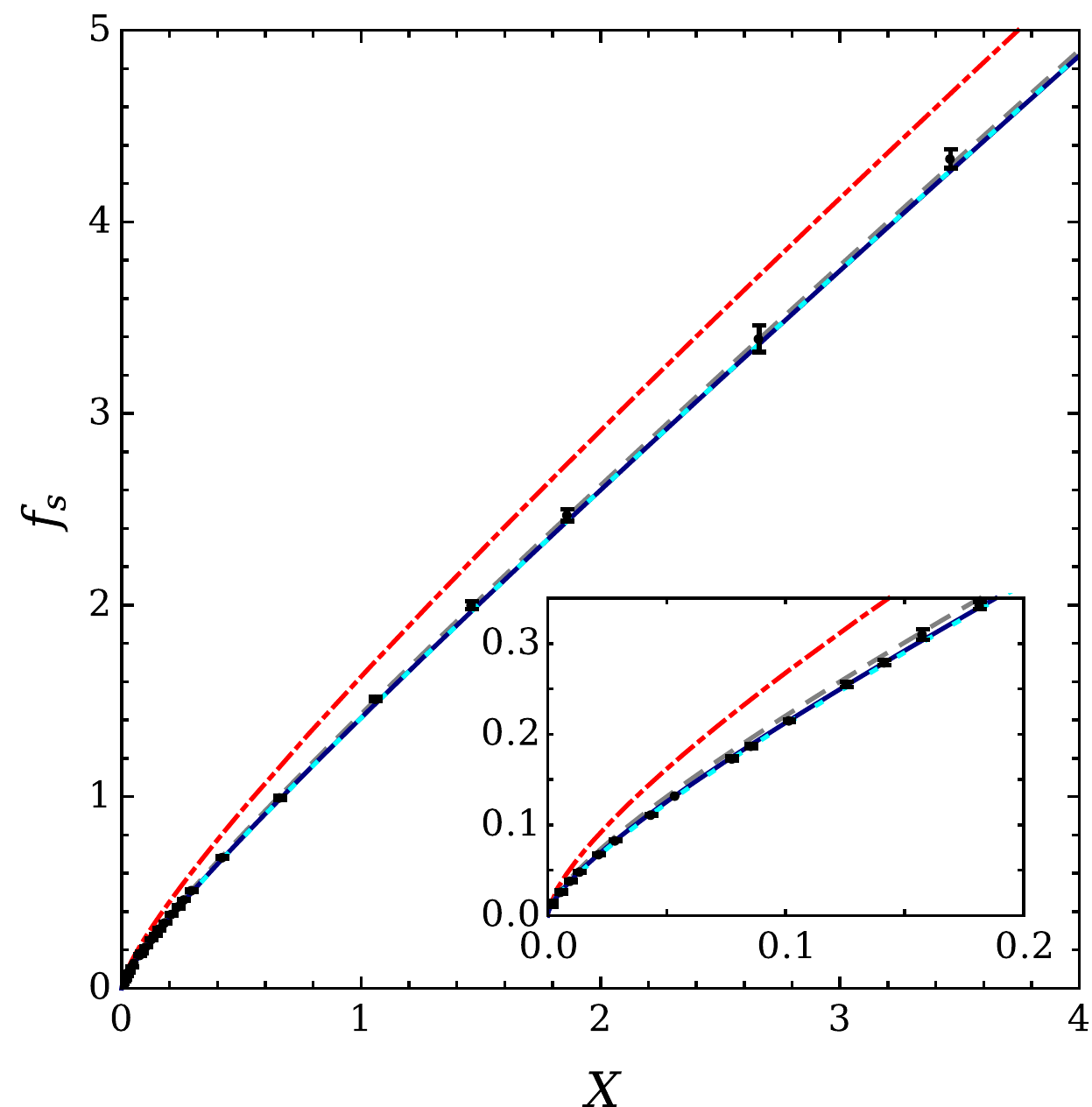}}
 \hspace{0.5cm}
 \subfloat[]{\includegraphics[width=.48\textwidth]{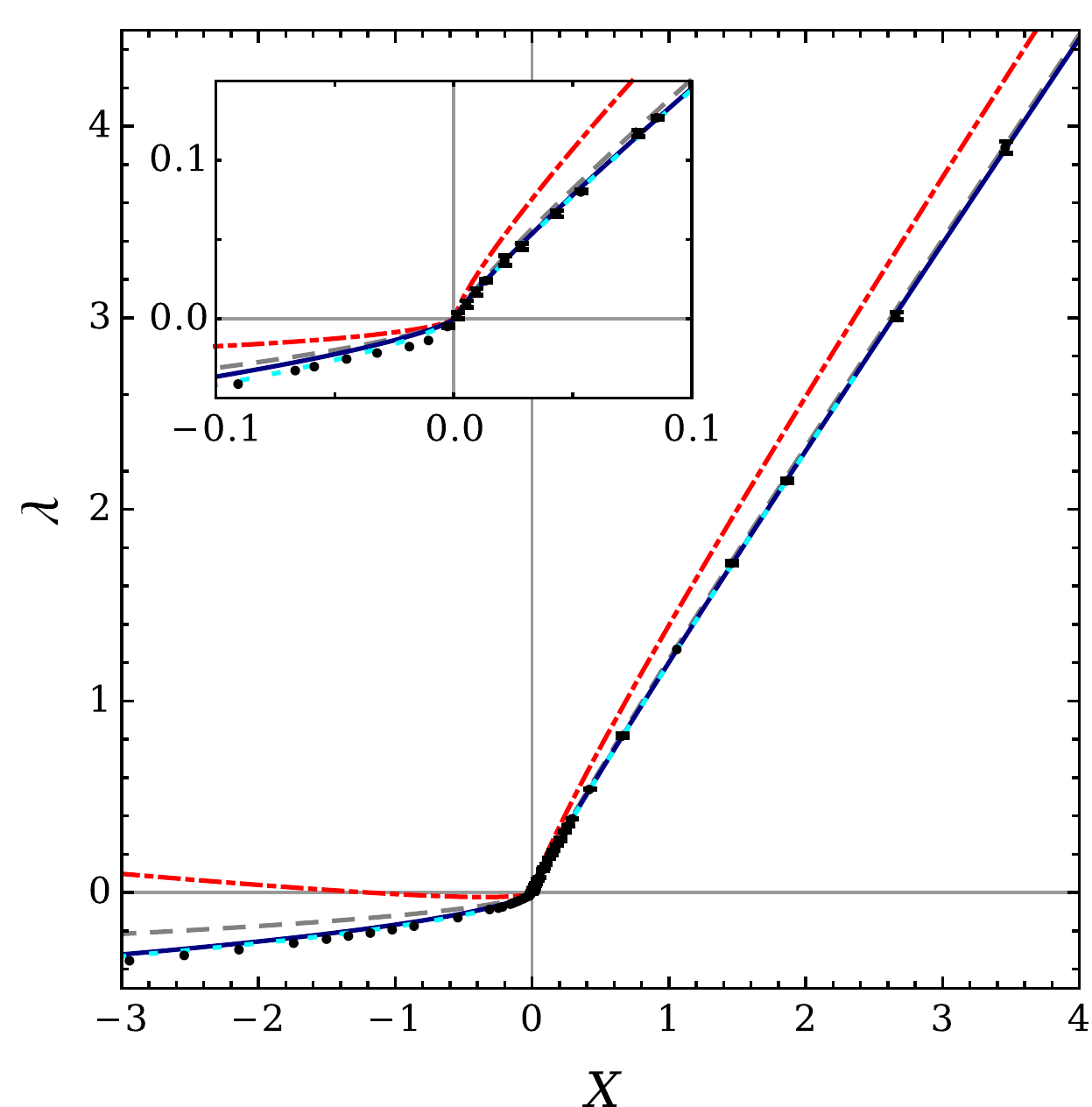}}
\caption{\label{fig:fs_lambda_3D_nu} Dimensionless superfluid density $f_s$ (a)  and density profile $\lambda$ (b) as a function of $X$ in three dimensions.
The red dash-dotted lines are obtained using the interpolating basis with $\nu=1$, the gray dashed lines with $\nu=1.5$ and the dark blue solid lines with $\nu=2$.
The cyan dotted lines are obtained using the Cartesian representation.
 The black circles correspond to the MC simulations of Ref.\ \cite{Prokofev2004}. The insets illustrate details of the behavior near $X=0$.}
\end{figure}

Fig.\ \ref{fig:fs_lambda_3D_nu} shows our results for the dimensionless functions $f_s(X)$ and $\lambda(X)$ for different choices of the parameter $\nu$ in Eq. (\ref{eq:bk}), as well as the results obtained with the Cartesian representation. We can see that by increasing the value of $\nu$, that is by making the transition between the two representations more abrupt, the curves for both $f_s$ and $\lambda$ get closer to the results in the Cartesian representation giving a better agreement with the MC simulations. Our interpretation is that for small values of $\nu$ there are small admixtures of the AP representation in the Gaussian regime that result in a incorrect flow in the UV. In addition, for cases where the flow reaches the symmetric phase, this always occurs in the Gaussian regime and so a large enough $\nu$ ensures that we use the Cartesian representation during the entire flow.
Our calculations suggest that for $\nu\geq 2$ the curves converge. 

\subsection{Two dimensions}

\begin{figure}[t!]
\centering
\includegraphics[width=.5\textwidth]{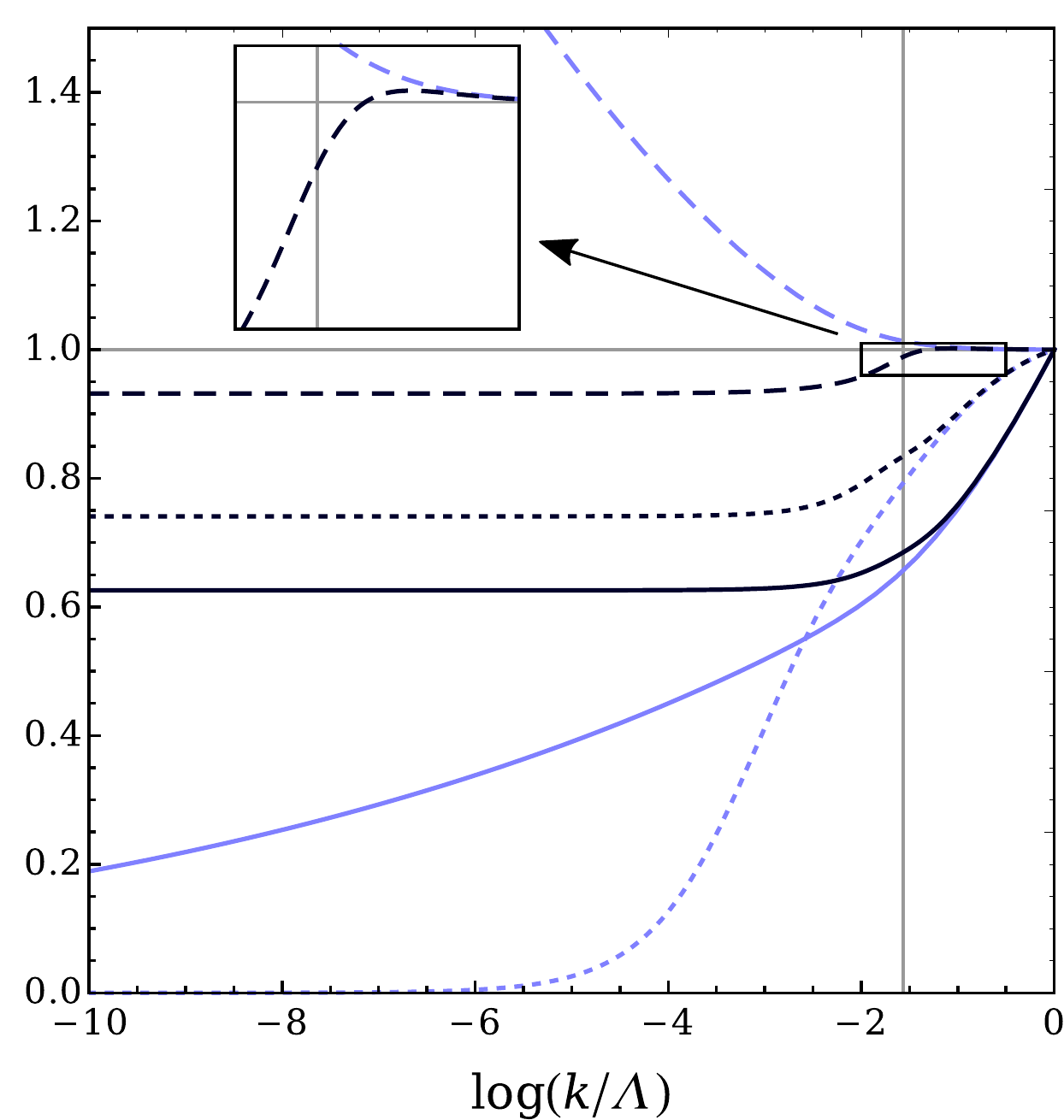}
\caption{\label{fig:flow2D} Flow of $\rho_0$ (solid lines), $u_2$ (dotted lines) and $Z_\ctheta$ (dashed lines) as a function of $\log(k/\Lambda)$ in two dimensions for $\mu_0=2\times 10^{-2}$.  The values of $\rhok0$ and $\uk2$ have been scaled by their initial values at $k=\Lambda$. The light blue lines are the flow obtained using the Cartesian representation  and the dark blue lines are obtained using the interpolating basis with $\nu=3$. The vertical line corresponds to the scale $k_h$ defined in Eq.\ (\ref{eq:kh}). }
\end{figure}

Fig.\ \ref{fig:flow2D} shows a typical evolution in the superfluid phase in two dimensions. Again, in the Gaussian regime all the flows coincide, but they diverge for smaller $k$. Unlike what is seen in three dimensions, the evolution of $\rho_0$ is very different for the two representations. The flow of $\rho_0$ in the Cartesian representation, where it should be interpreted as the condensate density $\rho_c$, decays approximately as a power law below $k_h$. However, $\rho_0$ flows to zero at a small but finite value of $k$ ($\log(k/\Lambda)\approx-50$ in this case, outside the range of the figure), implying that the symmetry is restored.   With the interpolating basis on the other hand, $\rho_0$ quickly converges to a finite value, showing that the system has a finite quasi-condensate density $\rho_q$. 

As in three dimensions, $u_2$ behaves quite differently in the two representations. In the Cartesian representation it flows to zero, vanishing quadratrically with $k$ in this case. In the interpolating basis, it converges quickly to a finite value. Since both $\rho_0$ and $u_2$ are finite in the physical limit, the amplitude fluctuations remain gapped in the interpolating basis. As in three dimensions, the evolution of $Z_\ctheta$ in the interpolating representation is consistent with $\rho_0$ changing from $\rho_c$ to $\rho_q$, initially growing larger than one before converging to a value below one. Since both $\rho_0$ and $Z_\ctheta$ are finite in the physical limit, this representation leads to a finite physical superfluid density. In contrast, $Z_\ctheta$ remains larger than one in the Cartesian representation.

\begin{figure}[t!]
\centering
\subfloat[]{\includegraphics[width=.48\textwidth]{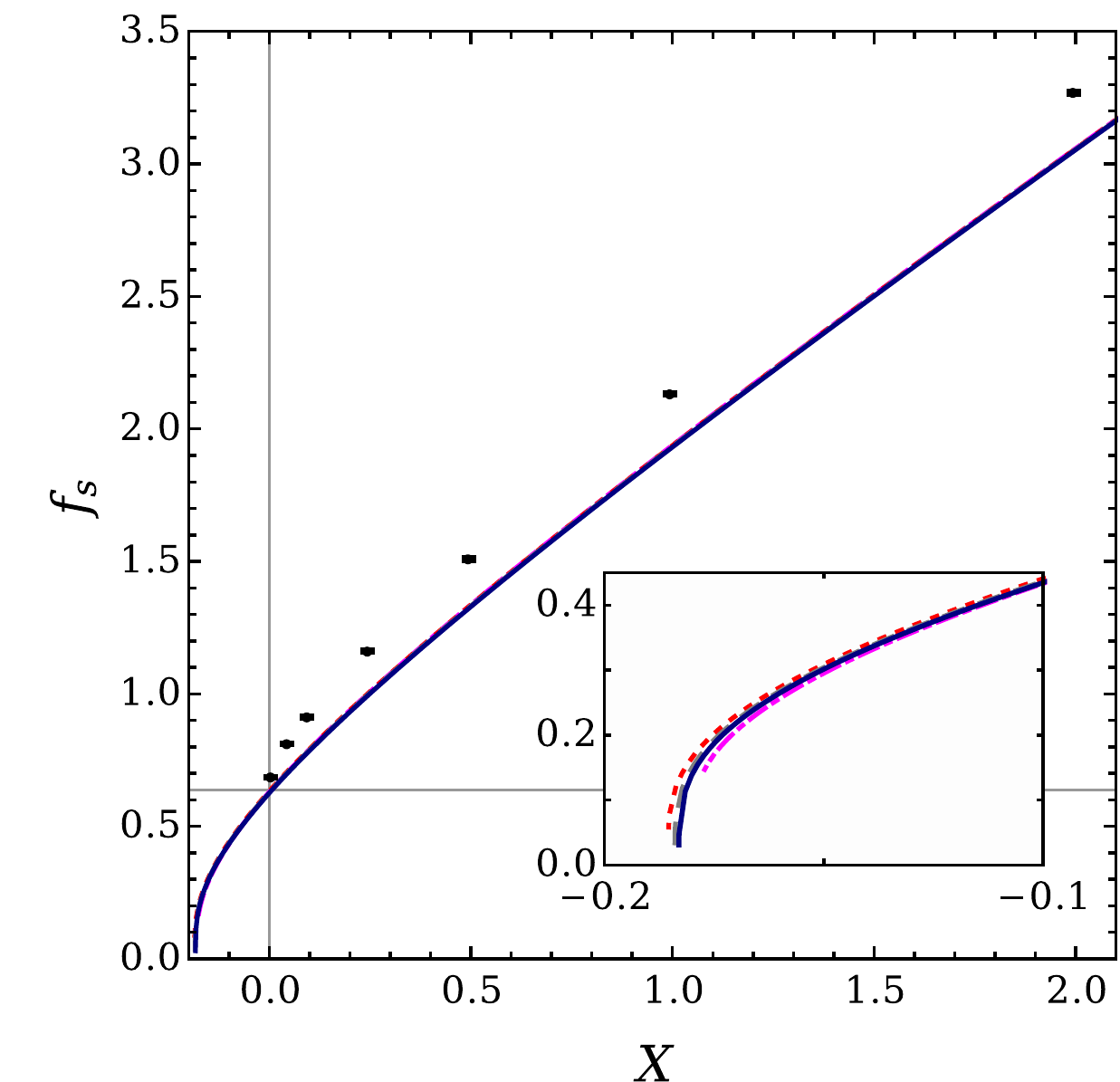}}
\hspace{0.5cm}
\subfloat[]{\includegraphics[width=.48\textwidth]{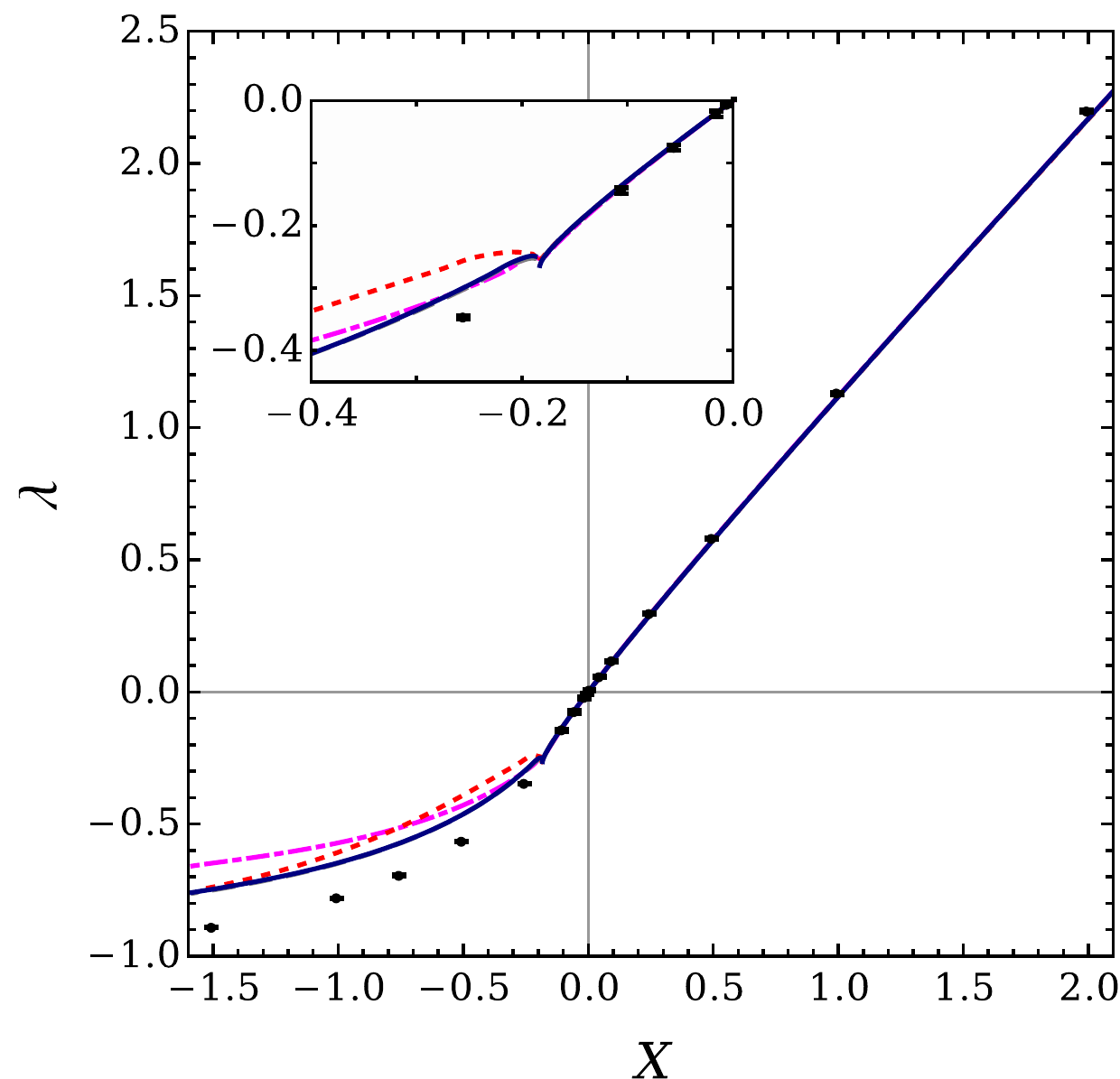}}
\caption{\label{fig:fs_lambda_2D_nu}  Dimensionless superfluid density $f_s$ (a) and density profile $\lambda$ (b) as a function of $X$ in two dimensions. 
The point where $X=0$ in the FRG calculations was chosen when $f_s=2/\pi$. 
The fuchsia dash-dotted lines are obtained using $\nu=1.5$, the red dotted lines using $\nu=2$, the gray dashed lines using $\nu=2.5$ and the
solid dark blue lines using $\nu=3$.
The black circles correspond to the MC simulations of Ref.\  \cite{Prokofev2002}.}
\end{figure}

Fig.\ \ref{fig:fs_lambda_2D_nu} shows results for $f_s(X)$ and $\lambda(X)$ for different choices of $\nu$. The phase transition in our approach is different from the true BKT transition. To make it easier to compare with the results of simulations, we have shifted our definition of $X$ so that $X=0$ corresponds to $f_s=2/\pi$ (the known value of the superfluid density at the BKT transition). Without this adjustment the normal phase is not reached in our calculations until $X\approx-0.18$. We show results only for $\nu\geq 1.5$; for smaller values of $\nu$ the flow is always driven to the symmetric phase. Our interpretation is that for small values of $\nu$   there are admixtures of the Cartesian representation in the Goldstone regime that make the flow unstable.

For $\nu\geq 1.5$, both $f_s$ and $\lambda$ are rather insensitive to the choice of $\nu$ in the region $X\gtrsim -1$.  On the other hand,
we see a more noticeable dependence of $\lambda(X)$ on the value of $\nu$ in the normal phase ($X\lesssim-0.2$). In all cases, the results depend only weakly on $\nu$ for $\nu \geq 2.5$.  Around the superfluid-to-normal phase transition ($-0.2\lesssim X \lesssim -0.1$) we see the most noticeable differences in $f_s(X)$ as well as a discontinuity in $\lambda(X)$. 

We expect that for normal systems the flow should reach the symmetric phase in the Gaussian regime ($k\gg k_h$), and so the entire flow is solved using the Cartesian representation. This is always the case in three dimensions and in two dimensions for $X\lesssim-0.2$. However, close to the phase transition in two dimensions ($X\approx -0.18$) the flow reaches the symmetric phase for $k<k_h$ (the Goldstone regime), and thus we are using a mixed representation in a region where it is not applicable. This causes an instability around the phase transition, as well as the discontinuity in $\lambda(X)$. This behavior is probably an artifact caused by the neglect of vortex effects. In particular, we note that a similar discontinuity in $\lambda$ was reported by Lim \emph{et al.}~\cite{Lim2008} when using the AP representation of Ref.~\cite{AlKhawaja2002}.

In two dimensions, vortices are important for describing the superfluid phase close to the phase transition. In addition, the change of the vortex fugacity produces a jump in the superfluid density at $f_s=2/\pi$ \cite{Berezinskivi1971,Kosterlitz1973}.  Although there is a noticeable difference between our results for $f_s$ and the MC simulations in the superfluid phase ($X>0$), the inclusion of vortex effects should produce a decrease of the value of $f_s$ around the BKT transition.  

Vortices are also important for describing the normal phase where the Bose gas forms a vortex plasma.  
The omission of their effects could explains the deviations in the normal phase 
and the problems found around the phase transition ($X\approx -0.18$). In view of this, our results are in reasonable agreement with the simulations. The results for $f_s$ in the region $X>0$ show larger deviations, but they are in line with those recently reported by Defenu \emph{et al.} \cite{Defenu2017} who used the FRG in the AP representation. On the other hand, the results for $\lambda$, although not as robust as the ones reported in Ref.~\cite{Rancon2012}, are in better agreement with the simulations, particularly for $X>0$.

 The smooth decrease of $f_s$, as opposed to a sudden jump, is believed to be a consequence of neglecting vortex effects. Unlike the hydrodynamic description at mean-field level, which predicts a quasi-condensate at any finite temperature, we find this only for a limited range of temperatures. The transition occurs at a lower density than that expected for the BKT one, which is similar to the  estimate by Fisher and Hohenberg \cite{Fisher1988}.

\section{Conclusions}
\label{sec:Conclusions}

In this work, we use the FRG to study weakly interacting Bose gases in two and three dimensions. Previous applications of the method have yielded encouraging results for critical behavior and bulk thermodynamic properties in three dimensions, when using the standard, Cartesian representation of the fields. However, particularly in two dimensions, they encountered problems with the IR behavior, with flows that restore the symmetry at
finite scales for any nonzero temperature, and hence do not lead to a physical superfluid phase. 
The alternative AP representation alleviates these IR problems but cannot easily be matched in the UV onto the bare interaction between the particles.

Building on the suggestion of Lamprecht and Pawlowski, we use a scale-dependent representation of the boson fields that interpolates smoothly between the Cartesian and AP representations. With this, we can solve the FRG flow using the Cartesian representation in the high-momentum regime, where both longitudinal and Goldstone modes are important, and using the AP representation in the low-momentum regime, where the Goldstone (phase) fluctuations dominate. 
This also allows superfluid systems to be described in terms of a quasi-condensate density, which can be non-zero in the absence of long-range-order.

In the present study, we work within the classical-field approximation, that is neglecting the UV quantum fluctuations by not considering the time derivative in the action. In this approximation, the Bose gases are described by O(2) models. This allows us to explore their properties around the phase transition and to test the interpolation between representations by comparing our results for the superfluid and boson densities with Monte-Carlo simulations. 

We adopt here the version of the FRG based on the average effective action, the flow of which is governed the Wetterich equation. Because we employ scale-dependent fields, we use a modified version of this equation with additional terms arising from the derivatives of the fields. We truncate the effective action to a quartic potential and terms with two spatial derivatives. The derivative terms include two structures which can distinguish amplitude and phase fluctuations in the phase with broken symmetry. In this first exploration of the approach, we use the ``optimized'' regulator suggested by Litim, as this eases numerical computation. 

In both two and three dimensions, we are able to obtain consistent descriptions of superfluid phases in Bose gases. We find
reasonable agreement with Monte-Carlo simulations by using interpolators that switch between the Cartesian and AP representations around the healing scale $k_h$ defined in Eq.~(\ref{eq:kh}), and that make this transition sharply enough. Below this scale, the ungapped phase fluctuations become increasingly important, making the AP representation the natural one to use. On the other hand, for scales above $k_h$, Goldstone and longitudinal fluctuations are of similar importance and the system should be described with the Cartesian representation in order to match the flow onto the bare action. 

Choosing a switching scale too far above $k_h$ or making the transition too slowly generates admixtures of the AP representation in the UV regime. 
The derivative interactions between the Goldstone modes become strong here, leading to results that deviate from those of the simulations. On the other hand, a switching scale too far below $k_h$ leaves admixtures of the Cartesian representation in the IR regime and, in two dimensions, these drive the flow to the non-superfluid symmetric phase. 

In three dimensions, our results for zero-momentum properties such as the superfluid density are very similar to those from the Cartesian version. The values of the parameter $\rho_0$, the minimum of the potential, are similar, although its interpretation is quite different in the two representations. In the Cartesian case, it is the condensate density, while in the AP case, it is the quasi-condensate. The wave function renormalization factors are different, if only by a few percent, reflecting the fact that the superfluid density, Eq.~(\ref{eq:scd}), should lie between the condensate and quasi-condensate. In contrast, the interaction strength $u_2$ is very different. This vanishes in the physical limit for the Cartesian case, as required by the Ward identity for the longitudinal self-energy, but it remains finite for the AP case, where it describes the interaction between amplitude modes. Those modes are fluctuations of an invariant field and so are not constrained by the symmetry.

In two dimensions, the condensate density at finite temperature vanishes in the physical limit, as required by the Coleman-Mermin-Wagner theorem. By switching to the AP representation, the interpolating fields allow us to work with a quasi-condensate density instead. This is finite in two dimensions and gives gapped amplitude modes. As a result we obtain a stable superfluid phase at finite temperature, in contrast to previous FRG studies that used the Cartesian representation. We find at least qualitative agreement between our results and Monte-Carlo simulations. However, there are also important differences. Most notably, although we do see a phase transition between superfluid and normal phases, we do not reproduce the features of BKT transition. Moreover, we find numerical instabilities around this transition and there are noticeable deviations in the values of the boson density in the normal phase. These are not unexpected because our current treatment does not include the vortex effects.

This inability to reproduce BKT physics is a weakness of our current approach. Nonetheless, the use of the AP representation makes it much easier to describe the IR regime, with a simple truncation of the derivative expansion giving reasonable results in the physical limit. In addition, 
it should be possible to extend the approach to implement vortex effects by taking into account the periodicity of the phase fields in this representation. This in contrast to the Cartesian version of the FRG which,  even though it reproduces some aspects of the BKT transition, does not include vortex physics in a controlled way.

Having demonstrated the usefulness of this approach for the simpler case of Bose gases in the classical-field approximation, the next step will be to apply it to fully dynamical systems, that is, keeping terms with time derivatives in the action. This will allow us to study these gases from zero temperature through the superfluid phase transition, and even to apply the FRG to one-dimensional systems. We expect that, by correctly treating the low-momentum phase fluctuations, the interpolation should have a greater impact at lower temperatures, solving the IR issues that have been encountered with the FRG in the Cartesian representation. We also plan to extend the approach to study a more complete set of thermodynamic variables, such as the pressure and the entropy, and to examine more general choices of regulator. Finally, in order to make contact with the physics of the BKT transition, we intend to explore ways to include vortices within an FRG treatment.

\begin{acknowledgments}
FI acknowledges funding from CONICYT Becas Chile under Contract No 72170497.
MCB and NRW are supported by the UK STFC under grants ST/L005794/1 and ST/P004423/1.
We acknowledge helpful discussions with Jan Pawlowski and Henk Stoof. We thank Jan Pawlowski for providing us with Ref.~\cite{Lamprecht2007}.
\end{acknowledgments}

\appendix

\section{Symmetric phase}
\label{app:SymmetricPhase}

Although the symmetric phase is not the focus of this work, we do find a transition to the symmetric phase for some calculations  close to the critical point. In the symmetric phase we work with the original conjugate representation of the fields ($\phi$ and $\phid$), then we use Ansatz (\ref{eq:Gammak}) imposing $\rho_0=0$ in the effective potential. The propagator evaluated at $\phi=\phid=0$ is given by
\begin{align}
G_{\phi\phid}(\Q)=&G_{\phid\phi}(\Q) \nonumber\\
=&\frac{-1}{Z_m\Q^2/2m+u_1-n_1\delta\mu+R(\Q)},
\end{align} 
and the off-diagonal terms $G_{\phi\phi}(\Q)=G_{\phid\phid}(\Q)=0$. The  regulator takes the form
\begin{equation}
R(\Q)=\frac{Z_m}{2m}(k^2-\Q^2)\theta(k^2-\Q^2).
\end{equation}

\begin{figure}[t!]
    \subfloat[]{\includegraphics[scale=1.5]{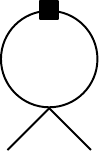}}
    \hspace{1cm}
    \subfloat[]{\includegraphics[scale=1.5]{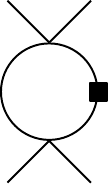}}
    \caption{\label{fig:DiagramsSP} Diagrammatic representation of $\vGamma^{(2)}$ (a) and $\vGamma^{(4)}$ (b) in the symmetric phase. The lines denote the propagator and the filled squares denote $\partial_k \bm{R}$.}    
\end{figure}

Since the fields $\phi$ and $\phid$ are $k$-independent, the flow equations for the running couplings are obtained from taking functional derivatives of Eq.\ (\ref{eq:WetterichEq}). They are
\begin{align}
\vu1 =& -\vGamma^{(2)}_{{\phi}{\phid}}\Big|_{\rho_0,\mu_0},\nonumber\\
2\vu2 =&-\vGamma^{(4)}_{{\phi}{\phi}{\phid}{\phid}}\Big|_{\rho_0,\mu_0},\nonumber \\
\vn0 =&\,\partial_{\mu}\vGamma\Big|_{\rho_0,\mu_0}, \nonumber \\
\vn1 =& \,\partial_{\mu}\left(\vGamma^{(2)}_{{\phi}{\phid}}\right)\Big|_{\rho_0,\mu_0}, \nonumber\\
2\vn2 =& \,\partial_{\mu}\left(\vGamma^{(4)}_{{\phi}{\phi}{\phid}{\phid}}\right)\Big|_{\rho_0,\mu_0}, \nonumber\\
\frac{\vZm}{m} =& -2\partial_{\vP^2}\left(\vGamma^{(2)}_{{\phi}{\phid}}\right)\Big|_{\rho_0,\mu_0,\vP=0},\nonumber\\
\frac{\vYm}{m} =& -4\partial_{\vP^2}\left(\vGamma^{(4)}_{{\phi}{\phi}{\phid}{\phid}}\right)\Big|_{\rho_0,\mu_0,\vP=0},
\end{align}
where everything is evaluated at $\phi=\phid=0$ and $\mu=\mu_0$. The diagrammatic representation of the flow equations is shown in Fig.\ \ref{fig:DiagramsSP}, where most diagrams vanished because $\phi=\phid=0$. The explicit expressions for these diagrams is given in Appendix \ref{app:DiagramsBP}. In order to isolate the evolution of the mass renormalizations, for $\vZm$ the external legs carry momentum $\vP$, while for $\vYm$ they carry momentum $\vP/2$. 

\section{Explicit expressions of the diagrams}
\label{app:DiagramsBP}
In the following we present the explicit form which take the different diagrams in both the broken and symmetric phase in terms of the matrices involved. As it was mentioned in the main text, the particular expressions for the matrices have been generated with \texttt{Wolfram Mathematica}. Additionally, with our choice of regulator (\ref{eq:Rlitim}) the momentum integrals are solved analytically before numerically solve the flow equations. 

Here we note that the functional derivatives are evaluated in momentum space using the convention 
$\phi(\Q)=T\int\frac{\ddQ}{(2\pi)^d}e^{i \Q\cdot\X}\phi(\X)$.
\subsection{Broken phase}
The explicit expression for $\vGamma^{(1)}_{a}$ (diagrams in Fig. \ref{fig:FlowGamma1_b}) is
\begin{widetext}
\begin{align}
\vGamma^{(1)}_{a}=&\int_\Q\bigg(-\frac{1}{2}\partial_k \MR(\Q) \MG(\Q) \MGamma^{(3)}_a(\bm{0},\Q,-\Q) \MG(\Q)+\MR(\Q)\MG(\Q)\vPHI^{(2)}_a(\bm{0},\Q,-\Q)\nonumber\\
&-\MR(\Q)\MG(\Q)\MGamma^{(3)}_a(\bm{0},\Q,-\Q)\MG(\Q)\vPHI^{(1)}(\bm{0},\Q,-\Q)\bigg),
\label{eq:FlowGamma1_b}
\end{align}
and for $\vGamma^{(2)}_{ab}$ (diagrams in Fig. \ref{fig:FlowGamma2_b}) is
\begin{align}
\vGamma^{(2)}_{ab}=&\int_\Q\bigg(-\frac{1}{2}\partial_k\MR(\Q)\MG(\Q)\MGamma^{(4)}_{ab}(\vP,-\vP,\Q,-\Q)\nonumber\\
&+\frac{1}{2}\partial_k\MR(\Q)\MG(\Q)\MGamma^{(3)}_a(\vP,\Q,-\vP-\Q)\MG(\vP+\Q)\MGamma^{(3)}_b(-\vP,\vP+\Q,-\Q)\MG(\Q)\nonumber\\
&+\MR(\Q)\MG(\Q)\vPHI^{(3)}_{ab}(\vP,-\vP,\Q,-\Q)\nonumber\\
&-\MR(\Q)\MG(\Q)\MGamma^{(3)}_a(\vP,\Q,-\vP-\Q)\MG(\vP+\Q)\vPHI^{(2)}_b(-\vP,\vP+\Q,-\Q)\nonumber\\
&-\MR(\Q)\MG(\Q)\MGamma^{(3)}_b(-\vP,\Q,\vP-\Q)\MG(\Q-\vP)\vPHI^{(2)}_a(\vP,-\vP+\Q,-\Q)\nonumber\\
&-\MR(\Q)\MG(\Q)\MGamma^{(4)}_{ab}(\vP,-\vP,\Q,-\Q)\MG(\Q)\vPHI^{(1)}(\Q,-\Q)\nonumber\\
&+\MR(\Q)\MG(\Q)\MGamma^{(3)}_a(\vP,\Q,-\vP-\Q)\MG(\vP+\Q)\MGamma^{(3)}_b(-\vP,\vP+\Q,-\Q)\MG(\Q)\vPHI^{(1)}(\Q,-\Q)\nonumber\\
&+\MR(\Q)\MG(\Q)\MGamma^{(3)}_b(-\vP,\Q,\vP-\Q)\MGamma^{(3)}_a(\vP,-\vP+\Q,-\Q)\MG(\Q)\vPHI^{(1)}(\Q,-\Q)\bigg),
\label{eq:FlowGamma2_b}
\end{align}
\end{widetext}
where
\begin{equation}
\int_\Q=\int \frac{\ddQ}{(2\pi)^d},
\end{equation}
and the subscripts $a$ and $b$ denote a field $\sigma$ or $\ctheta$. The momentum $\vP$ correspond to the momentum added in the external legs in order to isolate the flow of the mass renormalizations, and it is always taken to zero after taking the derivatives. The matrices $\MGamma^{(n+2)}$ correspond to the vertices between $n$ external legs and a propagator, and are extracted from taking $n$ functional derivatives of the $2\times2$ matrix
\begin{equation}
\MGamma^{(2)}=\frac{\delta^2 \Gamma}{\delta \PHI^2},
\end{equation}
where $\Gamma$ is Ansatz (\ref{eq:Gammabk}). Similarly, the matrices $\vPHI^{(n+1)}$ correspond to the vertices between $n$ external legs and $\vPHI^{(1)}$ (black circles in the diagrams), and are extracted from taking $n$ functional derivatives of Eq. (\ref{eq:vPHI1int}). Note that although Eqs. (\ref{eq:FlowGamma1_b}) and (\ref{eq:FlowGamma2_b}) include the explicit dependency on momentum in all the matrices, the $\vPHI^{(n+1)}$ matrices are momentum-independent.

\subsection{Symmetric phase}
Following the notation used previously, the explicit expression for  $\vGamma^{(2)}$ in the symmetric phase (diagram in Fig \ref{fig:DiagramsSP}a) is
\begin{widetext}
\begin{equation}
\vGamma^{(2)}_{ab}=\int_\Q \partial_k\MR(\Q)\MG(\Q)\MGamma^{(4)}_{bb}\left(\frac{\vP}{2},-\frac{\vP}{2},\Q,-\Q\right)\MG(\Q)\MGamma^{(4)}_{aa}\left(\frac{\vP}{2},-\frac{\vP}{2},\Q,-\Q\right),
\end{equation}
and for $\vGamma^{(4)}$ (diagram in Fig (diagram in Fig \ref{fig:DiagramsSP}b) is
\begin{align}
\vGamma^{(4)}_{ab}=&\int_\Q \bigg(\partial_k\MR(\Q)\MG(\Q)\MGamma^{(4)}_{ab}\left(\frac{\vP}{2},-\frac{\vP}{2},\Q,-\Q\right)\MG(\Q)\MGamma^{(4)}_{ab}\left(-\frac{\vP}{2},\frac{\vP}{2},\Q,-\Q\right)\MG(\Q) \nonumber\\
&+\partial_k\MR(\Q)\MG(\Q)\MGamma^{(4)}_{ab}\left(\frac{\vP}{2},-\frac{\vP}{2},\Q,-\vP-\Q\right)\MG(\vP+\Q)\MGamma^{(4)}_{ab}\left(-\frac{\vP}{2},-\frac{\vP}{2},\vP+\Q,-\Q\right)\MG(\Q) \nonumber\\
&\partial_k\MR(\Q)\MG(\Q)\MGamma^{(4)}_{bb}\left(\frac{\vP}{2},-\frac{\vP}{2},\Q,-\Q\right)\MG(\Q)\MGamma^{(4)}_{aa}\left(\frac{\vP}{2},-\frac{\vP}{2},\Q,-\Q\right)\MG(\Q).
\end{align}
\end{widetext}

\section{Dependence on switching scale}
\label{app:alpha}

\begin{figure}[t!]
\centering
\includegraphics[width=.48\textwidth]{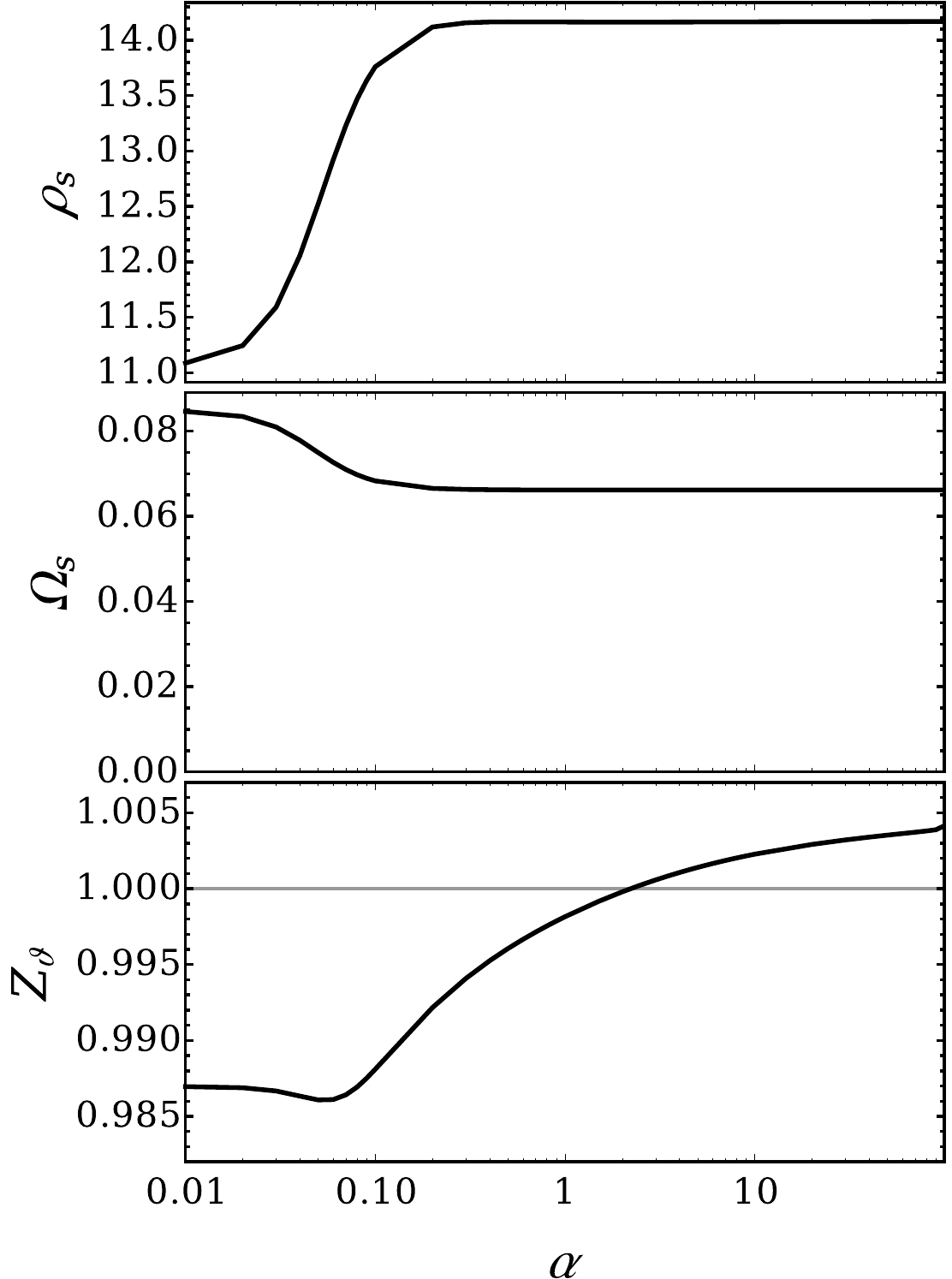}
\caption{\label{fig:alphas_3D} Physical superfluid density $\rho_s$, superfluid fraction $\Omega_s$ and phase renormalization $Z_\ctheta$ as a function of $\alpha$ in three dimensions for $\mu_0=1.5\times10^{-2}$. We use $\nu=3$, $T=10$, $g_0=10^{-3}$ and $\Lambda=1$. }
\end{figure}

Here we examine the dependence of our results on the parameter $\alpha$ in Eq.\ (\ref{eq:bk}), which allows us to vary the scale at which our interpolator switches between representations. In Fig.\ \ref{fig:alphas_3D} we show results for the observable  superfluid density $\rho_s$, the superfluid fraction $\Omega_s=\rho_s/n_0$ and the (not directly observable) phase renormalization factor $Z_\ctheta$, as functions of $\alpha$ in three dimensions. We use  the same initial conditions for all the calculations. Large values of $\alpha$ correspond to switching to the AP representation at scales well below $k_h$; we find that our results for both $\rho_s$ and $\Omega_s$ converge to the Cartesian ones for values of $\alpha \gtrsim 0.4$.  On the other hand, the behavior of $Z_\ctheta$ is more sensitive to the choice of $\alpha$.  Indeed, although the values of both $Z_\ctheta$ and $\rho_0$ show appreciable dependences on $\alpha$, in the  limit $k=0$ they lead to the same results for $\rho_s$ for $\alpha\geq 0.4$. 
For  $\alpha>2$, the value of $Z_\ctheta$ is greater than one. This would imply that the superfluid density $\rho_s$ is greater than the quasi-condensate density $\rho_q$, which is not physically correct. Thus, a safe choice for the value of $\alpha$ lies between 0.4 and 2.0.

\begin{figure}[t!]
\centering
\includegraphics[width=.48\textwidth]{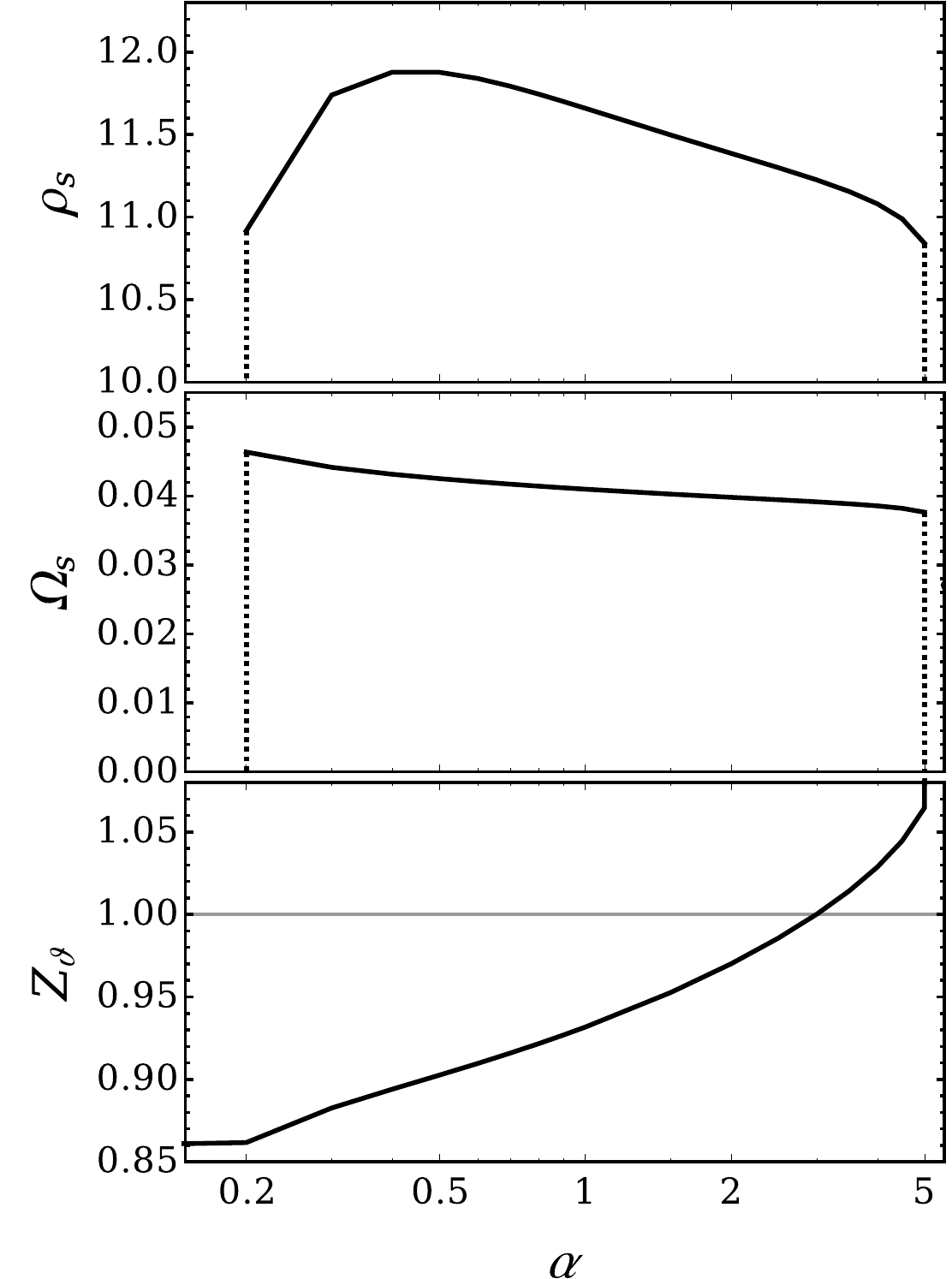}
\caption{\label{fig:alphas_2D} Physical superfluid density $\rho_s$, superfluid fraction $\Omega_s$ and phase renormalization $Z_\ctheta$  as a function of $\alpha$ in two dimensions for $\mu_0=2\times10^{-2}$.  We use $\nu=3$, $T=10$, $g_0=10^{-3}$ and $\Lambda=1$.}
\end{figure}

In Fig.\ \ref{fig:alphas_2D}  the same quantities are displayed in two dimensions. Note that the range of $\alpha$ shown is smaller than in three dimensions:
For $\alpha>5$, the dynamical part of the flow takes place entirely in the Cartesian representation driving $\rho_0$ to zero at a finite scale. Similarly, the flow is driven to the symmetric phase for $\alpha<0.2$ for the  choice of $\mu_0$ used,  $\mu_0=1.5\times10^{-2}$. We find that for $\alpha\ll 1$ the flow remains in the broken phase only for  unrealistically large values of $\mu_0$. The superfluid density is slightly more sensitive to $\alpha$ than in three dimensions. However, the results shown are near the BKT phase transition, where we know our results are not accurate; the differences become smaller deep in the superfluid phase where vortex effects are less important. Nonetheless, for $\alpha$ between 0.4 and 2.0, the variation in $\rho_s$ does not exceed  4\% with respect to its value at $\alpha=1$. In addition, the superfluid fraction, which is more relevant for the universal behavior, shows a similar variation over the range plotted. Lastly, as in  three dimensions, we find that $Z_\ctheta(k=0)$ is greater that one for $\alpha>2$. Thus we again conclude that a safe choice of $\alpha$ lies between 0.4 and 2.0.

\bibliography{Biblio}

\end{document}